\documentclass[10pt,journal]{IEEEtran} 
\def\loadbreqn{}

\usepackage[T1]{fontenc}
\usepackage{amsmath,amssymb,amsfonts,mathrsfs,bm}
\usepackage{amsthm}
\usepackage{cite}
\usepackage{array}
\usepackage[shortlabels]{enumitem}
\usepackage{graphicx}
\usepackage{epstopdf}
\usepackage{url}
\usepackage{color}
\usepackage{multirow}
\usepackage[table]{xcolor}
\usepackage[normalem]{ulem}
\usepackage{array}
\newcolumntype{L}[1]{>{\raggedright\let\newline\\\arraybackslash\hspace{0pt}}m{#1}}
\newcolumntype{C}[1]{>{\centering\let\newline\\\arraybackslash\hspace{0pt}}m{#1}}
\newcolumntype{R}[1]{>{\raggedleft\let\newline\\\arraybackslash\hspace{0pt}}m{#1}}
\usepackage{subcaption}
\usepackage{xparse}



\usepackage{glossaries}
\newacronym{wrt}{w.r.t.}{with respect to}
\newacronym{iid}{i.i.d.}{independent and identically distributed}
\newacronym{MIMO}{MIMO}{mulitple-input multiple-output}
\newacronym{AOA}{AOA}{angle-of-arrival}
\newacronym{AOD}{AOD}{angle-of-departure}
\newacronym{LOS}{LOS}{line-of-sight}
\newacronym{NLOS}{NLOS}{non-line-of-sight}
\newacronym{TOA}{TOA}{time-of-arrival}
\newacronym{TDOA}{TDOA}{time-difference-of-arrival}
\newacronym{RSS}{RSS}{received signal strength}
\newacronym{GNSS}{GNSS}{Global Navigation Satellite System}

\usepackage{float}

\ifx\notloadhyperref\undefined
	\ifx\loadbibentry\undefined
		\usepackage[hidelinks,hypertexnames=false]{hyperref} 
	\else
		\usepackage{bibentry}
		\makeatletter\let\saved@bibitem\@bibitem\makeatother
		\usepackage[hidelinks,hypertexnames=false]{hyperref}
		\makeatletter\let\@bibitem\saved@bibitem\makeatother
	\fi
\else
	\relax
\fi

\usepackage[capitalize]{cleveref}
\crefname{equation}{}{}
\Crefname{equation}{}{}
\crefname{claim}{claim}{claims}
\crefname{step}{step}{steps}
\crefname{line}{line}{lines}
\crefname{dmath}{}{}
\crefname{dseries}{}{}
\crefname{dgroup}{}{}
\crefname{Theorem}{Theorem}{Theorems}
\crefname{Corollary}{Corollary}{Corollaries}
\crefname{Proposition}{Proposition}{Propositions}
\crefname{Lemma}{Lemma}{Lemmas}
\crefname{Definition}{Definition}{Definitions}
\crefname{Example}{Example}{Examples}
\crefname{Assumption}{Assumption}{Assumptions}
\crefname{Remark}{Remark}{Remarks}
\crefname{Rem}{Remark}{Remarks}
\crefname{remarks}{Remarks}{Remarks}
\crefname{Theorem_A}{Theorem}{Theorems}
\crefname{Corollary_A}{Corollary}{Corollaries}
\crefname{Proposition_A}{Proposition}{Propositions}
\crefname{Lemma_A}{Lemma}{Lemmas}
\crefname{Definition_A}{Definition}{Definitions}

\usepackage{algorithm,algorithmic}

\ifx\loadbreqn\undefined
	\relax
\else
	\usepackage{breqn} 
\fi

\ifx\useTheoremCounter\undefined
\newtheorem{Theorem}{Theorem}
\newtheorem{Corollary}{Corollary}
\newtheorem{Proposition}{Proposition}

\else

\fi

\theoremstyle{remark}

\newcommand{\calB}{\mathcal{B}}

\newcommand{\calF}{\mathcal{F}}

\newcommand{\calI}{\mathcal{I}}

\newcommand{\calL}{\mathcal{L}}

\newcommand{\calN}{\mathcal{N}}

\newcommand{\calS}{\mathcal{S}}

\newcommand{\bA}{\mathbf{A}}

\newcommand{\bD}{\mathbf{D}}

\newcommand{\bM}{\mathbf{M}}

\newcommand{\bs}{\mathbf{s}}

\newcommand{\bV}{\mathbf{V}}

\newcommand{\by}{\mathbf{y}}

\newcommand{\bz}{\mathbf{z}}

\DeclareSymbolFont{bsfletters}{OT1}{cmss}{bx}{n}
\DeclareSymbolFont{ssfletters}{OT1}{cmss}{m}{n}
\DeclareMathSymbol{\bsfGamma}{0}{bsfletters}{'000}
\DeclareMathSymbol{\ssfGamma}{0}{ssfletters}{'000}
\DeclareMathSymbol{\bsfDelta}{0}{bsfletters}{'001}
\DeclareMathSymbol{\ssfDelta}{0}{ssfletters}{'001}
\DeclareMathSymbol{\bsfTheta}{0}{bsfletters}{'002}
\DeclareMathSymbol{\ssfTheta}{0}{ssfletters}{'002}
\DeclareMathSymbol{\bsfLambda}{0}{bsfletters}{'003}
\DeclareMathSymbol{\ssfLambda}{0}{ssfletters}{'003}
\DeclareMathSymbol{\bsfXi}{0}{bsfletters}{'004}
\DeclareMathSymbol{\ssfXi}{0}{ssfletters}{'004}
\DeclareMathSymbol{\bsfPi}{0}{bsfletters}{'005}
\DeclareMathSymbol{\ssfPi}{0}{ssfletters}{'005}
\DeclareMathSymbol{\bsfSigma}{0}{bsfletters}{'006}
\DeclareMathSymbol{\ssfSigma}{0}{ssfletters}{'006}
\DeclareMathSymbol{\bsfUpsilon}{0}{bsfletters}{'007}
\DeclareMathSymbol{\ssfUpsilon}{0}{ssfletters}{'007}
\DeclareMathSymbol{\bsfPhi}{0}{bsfletters}{'010}
\DeclareMathSymbol{\ssfPhi}{0}{ssfletters}{'010}
\DeclareMathSymbol{\bsfPsi}{0}{bsfletters}{'011}
\DeclareMathSymbol{\ssfPsi}{0}{ssfletters}{'011}
\DeclareMathSymbol{\bsfOmega}{0}{bsfletters}{'012}
\DeclareMathSymbol{\ssfOmega}{0}{ssfletters}{'012}

\newcommand{\bbeta}{\bm{\beta}}
\newcommand{\bgamma}{\bm{\gamma}}

\newcommand{\btheta}{\bm{\theta}}
\newcommand{\bmu}{\bm{\mu}}
\newcommand{\bnu}{\bm{\nu}}

\newcommand{\bpi}{\bm{\pi}}

\newcommand{\bzeta}{\bm{\zeta}}
\newcommand{\bmeta}{\bm{\eta}}

\newcommand{\bphi}{\bm{\phi}}
\newcommand{\bpsi}{\bm{\psi}}
\newcommand{\bomega}{\bm{\omega}}
\newcommand{\bxi}{\bm{\xi}}
\newcommand{\blambda}{\bm{\lambda}}

\newcommand{\bOmega}{\bm{\Omega}}

\newcommand{\hgamma}{\widehat{\gamma}}

\DeclareMathOperator*{\argmax}{arg\,max}
\DeclareMathOperator*{\argmin}{arg\,min}

\newcommand{\qednew}{\nobreak \ifvmode \relax \else
      \ifdim\lastskip<1.5em \hskip-\lastskip
      \hskip1.5em plus0em minus0.5em \fi \nobreak
      \vrule height0.75em width0.5em depth0.25em\fi}

\newcommand{\tc}[1]{^{(#1)}}

\newcommand{\KLD}[2]{{D({#1}\ ||\ {#2})}}

\newcommand{\cond}[2]{\left. {#1}\, \middle| \, {#2} \right.}

\DeclareDocumentCommand \P { g d() g } {%
	\IfNoValueTF {#3} 
	{%
		\IfNoValueTF {#1} 
		{%
			\IfNoValueTF {#2}
			{%
				\mathbb{P}%
			}%
			{%
				\mathbb{P}\left({#2}\right)%
			}%
		}%
		{%
			\IfNoValueTF {#2}
			{%
				\mathbb{P}_{#1}%
			}%
			{%
				\mathbb{P}_{#1}\left({#2}\right)%
			}%
		}%
	}%
	{%
		\IfNoValueTF {#1} 
		{%
			\mathbb{P}\left(\cond{#2}{#3}\right)%
		}%
		{%
			\mathbb{P}_{#1}\left(\cond{#2}{#3}\right)%
		}%
	}%
}

\DeclareDocumentCommand \E { g o g } {%
	\IfNoValueTF {#3} 
	{%
		\IfNoValueTF {#1} 
		{%
			\IfNoValueTF {#2}
			{%
				\mathbb{E}%
			}%
			{%
				\mathbb{E}\left[{#2}\right]%
			}%
		}%
		{%
			\IfNoValueTF {#2}
			{%
				\mathbb{E}_{#1}%
			}%
			{%
				\mathbb{E}_{#1}\left[{#2}\right]%
			}%
		}%
	}%
	{%
		\IfNoValueTF {#1} 
		{%
			\mathbb{E}\left[\cond{#2}{#3}\right]%
		}%
		{%
			\mathbb{E}_{#1}\left[\cond{#2}{#3}\right]%
		}%
	}%
}

\newcommand{\Unif}[1]{\mathrm{Unif}\left(#1\right)}
\newcommand{\Dir}[1]{\mathrm{Dir}\left(#1\right)}
\newcommand{\Cat}[1]{\mathrm{Cat}\left(#1\right)}
\newcommand{\N}[2]{{\calN\left({#1},\ {#2}\right)}}
\newcommand{\Beta}[2]{{\calB e\left({#1},\ {#2}\right)}}

\definecolor{gray90}{gray}{0.9}

\newcommand{\msout}[1]{\text{\color{green} \sout{\ensuremath{#1}}}}
\newcommand{\del}[1]{{\color{green}\ifmmode \msout{#1}\else\sout{#1}\fi}}

\newcommand{\hide}[1]{}

\graphicspath{{./Figures/}} 
\newcommand{\tbomega}{\widetilde{\bomega}}

\newcommand{\LogNormal}[2]{{\mathrm{LogNormal}\left({#1},\ {#2}\right)}}

\ifCLASSOPTIONcompsoc
  \usepackage[nocompress]{cite}
\else
  \usepackage{cite}
\fi
\ifCLASSINFOpdf
\else
\fi
\begin{document}
%
\title{Using Social Network Information in Community-based Bayesian Truth Discovery}
%
%
%
%

\author{Jielong~Yang,~\IEEEmembership{Student Member,~IEEE,}
	Junshan~Wang,~\IEEEmembership{Member,~IEEE,}			
	and Wee~Peng~Tay,~\IEEEmembership{Senior~Member,~IEEE}
\IEEEcompsocitemizethanks{\IEEEcompsocthanksitem This research is supported by the Singapore Ministry of Education Academic Research Fund Tier 1 grant 2017-T1-001-059 (RG20/17).
\IEEEcompsocthanksitem J. Yang, J. Wang, and W. P. Tay are with the School of Electrical and Electronic Engineering, Nanyang Technological University, Singapore. \protect\\
Emails: jyang022@e.ntu.edu.sg, wangjs2@ntu.edu.sg, wptay@ntu.edu.sg.}

}

\IEEEtitleabstractindextext{%
\begin{abstract}
We investigate the problem of truth discovery based on opinions from multiple agents (who may be unreliable or biased) that form a social network. We consider the case where agents' reliabilities or biases are correlated if they belong to the same community, which defines a group of agents with similar opinions regarding a particular event. An agent can belong to different communities for different events, and these communities are unknown \emph{a priori}. We incorporate knowledge of the agents' social network in our truth discovery framework and develop Laplace variational inference methods to estimate agents' reliabilities, communities, and the event states. We also develop a stochastic variational inference method to scale our model to large social networks. Simulations and experiments on real data suggest that when observations are sparse, our proposed methods perform better than several other inference methods, including majority voting, TruthFinder, AccuSim, the Confidence-Aware Truth Discovery method, the Bayesian Classifier Combination (BCC) method, and the Community BCC method. 
\end{abstract}

}

\maketitle

\IEEEdisplaynontitleabstractindextext

%
\IEEEpeerreviewmaketitle

\section{Introduction}\label{sec:introduction}
In crowdsourcing and social sensing \cite{karger2013efficient,KanTay:C17,AceDahLobOzd:11,HoTayQue:J15,Tay:J15,GraSurAli2016,HuaWan2016,KanTay:J18}, information about the same event often comes from different agents. Agents may have their own biases and produce unreliable opinions. A commonly used approach to fuse the agents' opinions together is the majority voting method, which assumes that all agents have the same reliability \cite{LiGaoMen2015}. However, due to different backgrounds and access to prior information, agents' reliabilities or biases may vary widely. 

Truth discovery methods have been proposed to jointly estimate event truths and agent reliabilities by aggregating noisy information from multiple agents. For example, companies may gather product ratings from social media to estimate the popularity of their products \cite{MarWan2016} and regulatory agencies may use participatory social sensing to determine if certain events like traffic congestion have occurred by allowing the public to report such events to them \cite{GarXioSun2017}.  In this paper, we consider the truth discovery problem when social network information of the contributing agents is available. 

\subsection{Related Work}\label{subsec:relatedwork}
In \cite{WanKapLe2012,WanAmiLi2014,HuaWan2016,YaoHuLi2016}, probabilistic models are proposed for truth discovery from binary (true or false) observations. For binary observations, the reliability of each agent is the probability an event is true given that the agent reports it to be true. In \cite{WanKapLe2012},  maximum likelihood estimation (MLE) is used to estimate event truths and agent reliabilities. The Cramer-Rao lower bound (CRLB) on source reliability estimation is computed in \cite{WanKapAbd2013}. In these papers, agents are assumed to be independent of each other. 

A model proposed in \cite{WanAmiLi2014} assumes the dependency graphs of agents are disjoint trees. In \cite{HuaWan2016,YaoHuLi2016}, the dependency relationship is extended to general graphs and represented by a known dependency matrix. In \cite{MaTayXia:J18}, an agent can be influenced by another agent to change its observation to match that of the influencer. An iterative expectation maximization algorithm is developed to infer each agent's reliability and dependency on other agents.

Scalar reliabilities of agents are also used in truth discovery from multi-ary observations in \cite{YinHanPhi2008, DonBerSri2009, LiLiGao2015a, ZhaWanZha2017}.  In \cite{YinHanPhi2008}, the reliability of an agent is the probability that its opinion is correct. In this work, a Bayesian method named TruthFinder is proposed to iteratively estimate agent reliabilities and event truths.  In \cite{ DonBerSri2009}, the authors use the same definition of reliabilities as \cite{YinHanPhi2008} and develop a Bayesian method named AccuSim to detect copying relationship among agents and jointly infer agent dependencies, agent reliabilities, and event truths.  

In \cite{ LiLiGao2015a}, the reliability of an agent is regarded as the variance of the difference between the ground truth and the agent opinion and a Confidence-Aware Truth Discovery (CATD) method is proposed to deal with the phenomenon where most sources only provide a few claims. In the paper, the confidence interval of the reliability estimation is considered.  In \cite{ ZhaWanZha2017}, the reliability of an agent is the proportion of its opinions that are consistent with the ground truth of the events and the authors adopt a new constraint-aware Hidden Markov Model to effectively infer the evolving truths of events and the evolving reliabilities of agents. 

In \cite{ ZhaRubGem2012}, the reliability of each agent is represented with two parameters representing the false positive error and the false negative error respectively. In \cite{KimGha2012}, the authors further use a confusion matrix to represent the reliability of each agent, and proposes a method called Bayesian Classifier Combination (BCC) for truth discovery. In \cite{ZheLiLi2017}, the authors perform an evaluation on 17 algorithms using five datasets and draw the conclusion that the confusion matrix formulation generally performs better than scalar reliabilities, and BCC is among the best methods in decision making and single-label tasks. However, if each agent only observes a small subset of events, it is difficult to infer its reliability. In practice, agents having similar background, culture, socio-economic standings, and other factors, may form communities and share similar confusion matrices. The reference \cite{VenGuiKaz2014} proposed an extension of the BCC model, called the Community BCC (CBCC) model. In this model, the confusion matrix of an agent is a perturbation of the confusion matrix of its community. However, both \cite{KimGha2012} and \cite{VenGuiKaz2014} estimate the confusion matrices of agents based only on the agents' observations. The papers \cite{GraSurAli2016} and \cite{YinGraSur2016} show that agents in a crowd are not independent, but are instead connected through social ties, which can provide us with important information about which community an agent belongs to. In \cite{ZhaWuHua2017}, the authors consider the fact that the expertise (and thus the reliability) of each agent varies across events and use MLE to jointly estimate user expertise and event truths. 

\subsection{Main Contributions}
In this paper, we consider the use of social network information and community detection to aid in truth discovery based on agents' observations. Similar to \cite{VenGuiKaz2014}, we assume agents are clustered into communities for each observed event (an agent can belong to different communities for different events), and agents in the same community have similar confusion matrices. We use both the agents' observations and the social network connections among agents to jointly infer the communities and the event truths. 

Truth discovery on social networks often requires analyzing massive data. However, the traditional Gibbs sampling method used in \cite{KimGha2012} and variational inference method in \cite{VenGuiKaz2014} can not scale to a large dataset. The reason is that the entire dataset is used at each iteration of Gibbs sampling and variational inference, thus each iteration can be computationally expensive when the dataset is large. In our paper, we develop a three-level stochastic variational inference method for our Bayesian network model (see Chapter 3 of \cite{KolFri2009}) that can scale to large networks. The truths and communities are estimated iteratively. In each iteration, instead of re-analyzing the entire dataset (which includes information about the social network and agents' observations), we use randomly sampled sub-datasets to update the target variables. Our main contributions are as follows:
\begin{itemize}
	\item We propose a model that uses both social network information and agents' observations to jointly infer agent communities and event truths. To model the relationship between event states and the agents' observations, we use a mixed membership stochastic blockmodel \cite{AirBleFie2008} for the community structure and confusion matrices. Our model allows agents to switch communities when observing different events.	
	\item For small and medium sized networks, we develop a Laplace variational inference method to iteratively estimate the agent communities and event truths.
	\item We develop a three-level stochastic variational inference method for our model that can scale to large networks. 
	\item We perform simulations and experiments on real data, which demonstrate that our method performs better than majority voting, TruthFinder, AccuSim, CATD, BCC, and CBCC. 
\end{itemize}

The rest of this paper is organized as follows. In \cref{sec:Model&Notations}, we present our model and assumptions. In \cref{sec:VarInfer}, we develop a Laplace variational inference method for our Bayesian network model. In \cref{sec:StoVarInfer}, we develop a three-level stochastic variational inference method that can scale to large networks. Simulation and experiment results are presented in \cref{sec:SimulationResults}, and we conclude in Section \ref{sec:Conclusion}.

\emph{Notations:} We use boldfaced characters to represent vectors and matrices. Suppose that $\bA$ is a matrix, then $\bA(m,\cdot)$, $\bA(\cdot,m)$, and $\bA(m,n)$ denote its $m$-th row, $m$-th column, and $(m,n)$-th element, respectively. The vector $(x_1,\ldots,x_N)$ is abbreviated as $(x_i)_{i=1}^N$ or $(x_i)_i$ if the index set that $i$ runs over is clear from the context. We use $\Cat{p_1,\ldots,p_K}$, $\Dir{\frac{\alpha}{K_s},\ldots,\frac{\alpha}{K_s}}$, $\Unif{a,b}$, $\Unif{1,\ldots,R}$, $\Beta{g_0}{h_0}$ and $\N{\bM}{\bV}$ to represent the categorical distribution with category probabilities $p_1,\ldots,p_K$, the Dirichlet distribution with concentration parameters $\frac{\alpha}{K_s},\ldots,\frac{\alpha}{K_s}$, the uniform distribution over the interval $(a,b)$, the uniform distribution over the discrete set $\{1,\ldots,R\}$, the beta distribution with shape parameters $(g_0,h_0)$, and the normal distribution with mean $\bM$ and covariance $\bV$, respectively. We use $\Gamma(\cdot)$ and $\Psi(\cdot)$ to denote the gamma function and digamma function, respectively. The notation $\sim$ means equality in distribution. The notation $y\mid x$ denotes a random variable $y$ conditioned on $x$, and $p(y\mid x)$ denotes its conditional probability density function. $\E$ is the expectation operator and $\E_q$ is expectation with respect to the probability distribution $q$. We use $I(a,b)$ to denote the indicator function, which equals 1 if $a=b$ and 0 otherwise. We use $|\calS|$ to represent the cardinality of the set $\calS$.

\section{System Model}
\label{sec:Model&Notations}

\begin{figure}[!htb]
	\centering
	\includegraphics[width=0.95\columnwidth]{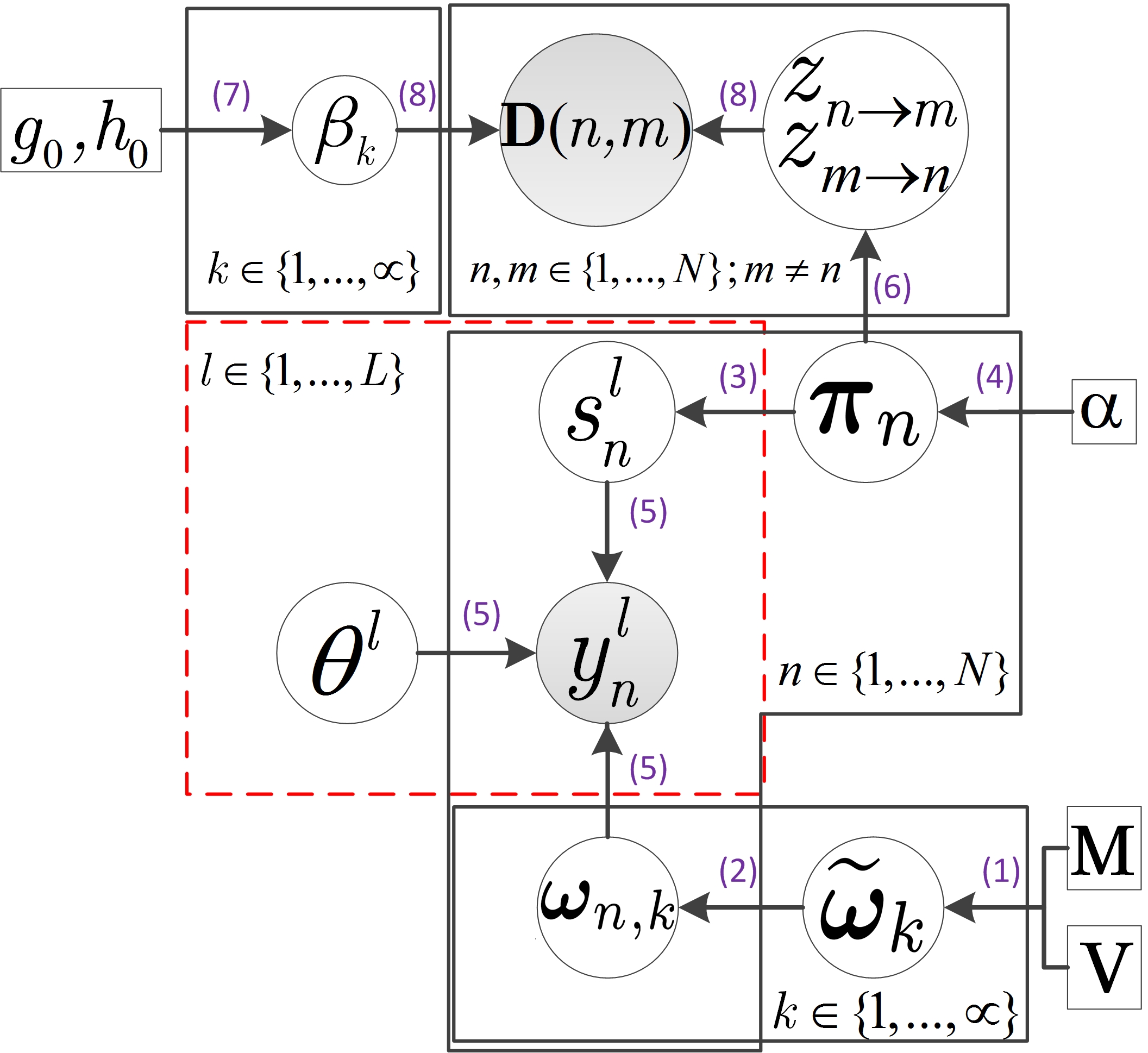}
	\caption{Our proposed Bayesian network model. Equation numbers are indicated on the corresponding edges.} 
	\label{fig:model}	
\end{figure}

In this section, we present our model and assumptions. Suppose that $N$ agents observe $L$ events and each event can be in $R$ possible states. Each agent observes only a subset of events, and provides its opinions of the events' states to a fusion center. The fusion center's goal is to infer the true state of each event from all the agents' opinions, and estimate the confusion matrix of each agent. 
The $(i,j)$-th element of the confusion matrix of agent $n$ is the probability that agent $n$'s opinion of an event with state $i$ is $j$.
We adopt the Bayesian network model shown in \cref{fig:model}, with notations used in the model summarized in \cref{table:Notations}. We explain the model in details below.  

\begin{table}[!htb]
	\caption{Summary of commonly-used notations.} 
	\centering 
	\begin{tabular}{L{1.5cm} L{4cm} C{2cm}} 
		\hline 
		Notation in \cref{fig:model} & Description & Variational Parameter in Section \ref{sec:VarInfer} \\ [0.5ex] 
		\hline 
		$\bD(n,m)=1\text{ (or 0)}$     &There is a (or no) social connection between agents $n$ and $m$. & N.A.\\ 
		\hline 
		$\bz=(z_{n\rightarrow m})_{n,m}$ & $z_{n\rightarrow m}$ is the index of the community agent $n$ subscribes to under the social influence of agent $m$.& $\bphi=((\phi_{n\rightarrow m,k})_k)_{n,m}$\\
		\hline
		$\bbeta=(\beta_k)_{k}$ & $\beta_k$ is the social network parameter defined in \eqref{eq:prior_Dnm}. & $\blambda=(\blambda_k)_{k}$; ${\blambda_k=(G_k,H_k)}$\\
		\hline
		$\bpi=(\bpi_{n})_n=(\pi_{n,k})_{n,k}$ & $\bpi_n$ is the distribution of $s_n^l$ and $z_{n\rightarrow m}$, which are defined in \eqref{eq:prior_s} and \eqref{eq:prior_znm}.& $\bgamma=(\gamma_{n,k})_{n,k}$\\
		\hline
		$\bs=(s_n^l)_{n,l}$ & $s_n^l$ is the index of the community whose belief agent $n$ subscribes to when it observes event $l$.& $\bpsi=((\psi_{n,k}^l)_k)_{n,l}$\\
		\hline
		$\by=(y_n^l)_{n,l}$ & $y_n^l$ is the observation of agent $n$ of event $l$.& N.A.\\
		\hline
		$\btheta=(\theta^l)_{l}$ & $\theta^l$ is the hidden true state of event $l$.& $\bnu=(\bnu^l)_{l}$ \\		
		\hline	
		$\tbomega=(\tbomega_{k})_{k}$   &  $\tbomega_k$ is the confusion matrix of community $k$.& $\bmu=(\bmu_{k})_{k}$ \\
		\hline	
		$\bomega=(\bomega_{n,k})_{n,k}$ & $\bomega_{n,k}$ is the confusion matrix of agent $n$ when it subscribes to the belief of community $k$. This is a perturbed version of $\tbomega_k$.& $\bxi=(\bxi_{n,s_n^l})_{n,l}$\\
		
		\hline
		$(\bM,\bV)$, $\alpha$, $(g_0,h_0)$   &  Known hyper-parameters defined in \eqref{eq: tomegaPrior}, \eqref{eq:OriginPiPrior}, and \eqref{eq:prior_znm}, respectively.& N.A.\\	
		\hline		
	\end{tabular}
	\label{table:Notations} 
\end{table}

We assume that a social network connecting the $N$ agents is known. Agents in a social network tend to form communities \cite{ForHri2016} whose members have similar interests or backgrounds. Agents in the same community may be more interested in certain events, and may share the same biases. An agent can subscribe to the beliefs of multiple communities. Consider an agent $n$ who observes an event $l$. Suppose that it decides to adopt the belief of community $k$ when observing event $l$. Let $\tbomega_k$ be the $R\times R$ confusion matrix of the community $k$ whose $r$-th row $\tbomega_{k}(r,\cdot)$ is assumed to follow a log-normal distribution modeled as:
\begin{align}
\tbomega_{k}(r,\cdot)\sim \LogNormal{\bM}{\bV},	\label{eq: tomegaPrior}	
\end{align}
where $\bM$ and $\bV$ are known hyper-parameters. The community confusion matrices $(\tbomega_k)_k$ are \gls{iid} for all $k$. Note that with the log-normal distribution assumption, $\tbomega_{k}(r,r')$ is positive. We suppose that the $R\times R$ confusion matrix $\bomega_{n,k}$ of agent $n$ is then given by drawing
\begin{align}
\bomega_{n,k}(r,\cdot)\sim \Dir{\tbomega_{k}(r,\cdot)}\label{eq: omegaPrior}
\end{align}
independently for each row $r$. The model \cref{eq: omegaPrior} allows us to correlate the individual confusion matrix $\bomega_{n,k}$ for every agent $n$ who subscribes to the belief of community $k$ through the community confusion matrix $\tbomega_{k}$.

Let $s_n^l$ denote the index of community whose belief agent $n$ subscribes to when it observes event $l$. We model it as
\begin{align}
s_n^l\mid\bpi_n \sim \Cat{ \bpi_n },\label{eq:prior_s}
\end{align}
where $\bpi_n\sim \mathrm{GEM}(\alpha)$ is \gls{iid} over $n$, and $\alpha$ is a concentration hyperparameter and GEM stands for the Griffiths, Engen and McCloskey stick breaking process \cite{TehJorBea2012}. Following \cite{IshZar2002,FoxSudJor2011}, we approximate the GEM process with its degree $K_s$ weak limit given by 
\begin{align}
\bpi_n \sim \Dir{\alpha/K_s, \ldots,\alpha/K_s}\label{eq:OriginPiPrior},
\end{align}   
where $K_s$ is the maximum number of communities. Let $\bpi_n = (\pi_{n,k})_{k=1}^{K_s}$.

Let $\theta^l$ be the hidden true state of event $l$ with prior distribution $\theta^l\sim \Unif{1, \ldots,R}$ and let $y_n^l$ denote the opinion of agent $n$ with respect to event $l$. We model the distribution of agent $n$'s opinion as:
\begin{align}
y_n^l\mid \theta^l,\{\bomega_{n,k}\}_{k=1}^{K_s},s_n^l\ \sim\ \Cat{\bomega_{n,s_n^l}(\theta^l,\cdot)}. \label{eq:Observation}
\end{align}
Our target is to estimate $(\theta^l)_{l=1}^L$ and $(\bomega_{n,k})_{n=1;k=1}^{N;K_s}$ from $(y_n^l)_{n=1;l=1}^{N;L}$. 

To model the available social network information, we suppose that the social network graph adjacency matrix $\bD$ is known, where $\bD(n,m)=1$ if agent $n$ and agent $m$ are connected, and $\bD(n,m)=0$ otherwise. We adopt the mixed membership stochastic blockmodel (MMSB)\cite{AirBleFie2008} to model $\bD(n,m)$. In this model, we use $z_{n\rightarrow m}$ to denote the community whose belief agent $n$ subscribes to due to the social influence from agent $m$. Under the influence of different agents, agent $n$ may subscribe to the beliefs of multiple communities. If both agents $n$ and $m$ subscribe to the belief of the same community, they are more likely to be connected in the social network. We assume the following:
\begin{align}
z_{n\rightarrow m}\mid \bpi_n& \sim \Cat{ \bpi_n },\nonumber\\
z_{m\rightarrow n}\mid \bpi_m& \sim \Cat{ \bpi_m },\label{eq:prior_znm} \\
\beta_{k}& \sim \Beta{g_0}{h_0}, \label{eq:prior_beta_k}
\end{align}
where $g_0, h_0 >0$ are hyperparameters and $k=1,\ldots,K_s$. We assume

\begin{align}
&\P(\bD(n,m)=1\mid z_{n\rightarrow m},z_{m\rightarrow n},\beta_{z_{n\rightarrow m}}) \nonumber\\
&=
\begin{cases}
\beta_{z_{n\rightarrow m}}, & \text{if}\ z_{n\rightarrow m}=z_{m\rightarrow n}, \\	
\epsilon, & \text{if}\ z_{n\rightarrow m}\ne z_{m\rightarrow n},
\end{cases}
\label{eq:prior_Dnm}
\end{align}
%
where $\epsilon$ is a small known constant. We also assume that $\{z_{n \rightarrow m}\}_{n,m}$ and $\{\beta_k\}_k$ are independent. Furthermore, conditioned on $(z_{n->m},z_{m->n})$, $\bD(m,n)$ is independent of $\bpi_n$. 

Combining \eqref{eq:prior_znm} and \eqref{eq:prior_Dnm}, we have
\begin{dmath}
p(z_{n \rightarrow m}=k\mid \bpi_n,z_{m \rightarrow n},\bD(n,m)=1,\beta_k)\\
\propto \beta_{k}^{I(z_{m \rightarrow n},k)}\epsilon^{(1-I(z_{m \rightarrow n},k))}\pi_{n,k},\label{eq:Posterior_z_1}
\end{dmath}
and
\begin{dmath}
p(z_{n \rightarrow m}=k\mid \bpi_n,z_{m \rightarrow n},\bD(n,m)=0,\beta_k)\\	
\propto (1-\beta_{k})^{I(z_{m \rightarrow n},k)}(1-\epsilon)^{(1-I(z_{m \rightarrow n},k))}\pi_{n,k}.\label{eq:Posterior_z_0}
\end{dmath}

\section{Laplace Variational Inference Method}\label{sec:VarInfer}
In this section, we present an approach to infer the  states of events and the confusion matrices of agents, based on our model. Let $\bbeta=(\beta_k)_{k}$, $\bz=(z_{n\rightarrow m})_{n,m}$, $\bs=(s_n^l)_{n,l}$, $\bpi=(\pi_{n,k})_{n,k}$, $\btheta=(\theta^l)_{l}$, $\bomega=(\bomega_{n,k})_{n,k}$, $\tbomega=(\tbomega_{k})_{k}$, and $\by=(y_n^l)_{n,l}$, where $n,m\in\{1, \ldots,N\}$, $m\neq n$, $k\in\{1, \ldots,K_s\}$, and $l\in\{1, \ldots,L\}$. For simplicity, let $\bOmega=\{\bbeta,\bz,\bs,\bpi,\btheta, \bomega,\tbomega\}$. As the closed-form of the posterior distribution $p(\bOmega\mid \by,\bD)$ is not available, we use a variational inference method \cite{HofBleWan2013} to approximate the posterior distribution.  The variational inference method first posits a family of densities, and then iteratively updates variational parameters to select a member in the family that has minimum Kullback Leibler (KL) divergence with the posterior distribution. Compared with Markov chain Monte Carlo (MCMC) sampling, variational inference methods solve an optimization problem and thus tends to be computationally faster. However, unlike MCMC, it does not guarantee that the global optimal \cite{BleKucMcA2017} inference is achieved. Variational inference methods are more suitable for large datasets, while MCMC is more appropriate for smaller ones. We describe our proposed variational method in detail below.

We use $\calF$ to denote a family of probability distributions over $\bOmega$. Our target is to find the member in $\calF$ that is closest to $p(\bOmega\mid \by,\bD)$ in KL divergence. We choose $\calF$ to be a mean-field variational family, so that the latent variables $\bbeta$, $\bz$, $\bs$, $\bpi$, $\btheta$, $\bomega$, $\tbomega$ are mutually independent and each is governed by a variational parameter. We further denote the variational parameters of the variational distribution of $\bbeta$, $\bz$, $\bs$, $\bpi$, $\btheta$, $\bomega$ and $\tbomega$ as $\blambda=(\blambda_k)_{k}$, $\bphi=((\phi_{n\rightarrow m,k})_k)_{n,m}$, $\bpsi=((\psi_{n,k}^l)_k)_{n,l}$, $\bgamma=(\gamma_{n,k})_{n,k}$, $\bnu=(\bnu^l)_{l}$, $\bxi=(\bxi_{n,k})_{n,k}$, and $\bmu=(\bmu_{k})_{k}$, respectively. Each set of parameters $\{\blambda,\bphi,\bpsi,\bgamma,\bnu,\bxi\}$ corresponds to a member in the mean-field variational family $\calF$. 
It is noteworthy that using the mean-field family is suboptimal. The mean-field family is expressive because it can capture any marginal density of the latent variables. However, it cannot capture correlations between them and thus will reduce the fidelity of the approximation and introduce a loss of optimality. Structured variational inference method can be used to introduce dependencies between the variables. However, the complexity of the family determines the complexity of the optimization. It is important to choose a family of distributions that contains a density close to the posterior distribution of the hidden variable, but simple enough for efficient optimization.
A distribution in $\calF$ is represented as
\begin{dmath}
q(\bOmega)=q(\bbeta;\blambda)q(\bz;\bphi)q(\bs;\bpsi)q(\bpi;\bgamma)q(\btheta;\bnu)q(\bomega;\bxi)q(\tbomega; \bmu)
=\prod_{k}q(\beta_k;\blambda_k)\prod_{(n,m):m \neq n}q(z_{n\rightarrow m};\bphi_{n\rightarrow m})\prod_{n,l}q(s_n^l;\bpsi_{n}^l)\\ \cdot \prod_n q(\bpi_{n};\bgamma_{n})\prod_l q(\theta^l;\bnu^l)\prod_{n,k}q(\bomega_{n,k};\bxi_{n,k})\prod_k q(\tbomega_{k};\bmu_{k}).\label{eq:Mean_Field}
\end{dmath}
To avoid cluttered notations, we omit the variational parameters for simplicity, e.g., we write $q(\bbeta)$ instead of $q(\bbeta;\blambda)$. 
Note that the variational density $q(\bOmega)$ is used to approximate the conditional density function $p(\bOmega\mid \by,\bD)$, i.e., all derivations and discussions throughout this section are conditioned on the observed $(\by, \bD)$.

In the variational inference method, we aim to find
\begin{align}\label{q*}
q^{*}(\bOmega)=\argmin_{q(\bOmega)\in\calF} \KLD{q(\bOmega)}{p(\bOmega\mid \by,\bD)},
\end{align}
where $\KLD{\cdot}{\cdot}$ is the KL divergence. From \cite{BleKucMcA2017}, finding \eqref{q*} is equivalent to maximizing the evidence lower bound
\begin{align}
\calL(q)\triangleq\E_{q(\bOmega)}[\log p(\bOmega,\by,\bD)]-\E_{q(\bOmega)}[\log q(\bOmega)]\label{eq:CostFunction}.
\end{align}

Since the prior distribution of $\tbomega$ and its likelihood in our model are not conjugate, we cannot use standard mean-field variational inference to update its variational parameter. Instead, we consider the Laplace variational inference approach proposed in \cite{WanBle2013}, which uses Laplace approximations \cite{TieKasKad1989} to approximate the variational distribution of $\tbomega$ with a normal distribution. For the other variables $\{\bbeta,\bz,\bs,\bpi,\btheta,\bomega\}$, the posterior distribution of each variable is in the same exponential family as its prior distribution and we choose the variational distribution of each variable from the same exponential family as its posterior distribution. Our target is to solve \eqref{eq:CostFunction} by updating these variational parameters iteratively. We call this the Variational Inference using Social network Information for Truth discovery (VISIT) algorithm, which is shown in \cref{alg:inferenceAlgorithm_VI}. In the following, we explain VISIT in detail by presenting our assumptions on the variational distribution of each variable in $\bOmega$, and deriving the procedure to iteratively update each variational parameter in our model.

\begin{algorithm}
	\caption{VISIT ($i$-th iteration)}\label{alg:inferenceAlgorithm_VI}
	\begin{algorithmic}[H]
		\renewcommand{\algorithmicrequire}{\textbf{Input:}}
		\renewcommand{\algorithmicensure}{\textbf{Output:}}
		\REQUIRE Variational parameters in $(i-1)$-th iteration, opinions $\by$, social network data $\bD$.
		\ENSURE  Variational parameters in $i$-th iteration. 		
		\FOR {each agent $n$ in $\{1, \ldots,N\}$}		
		\FOR {each agent pair $(n,m)$ in $\{(n,m)\}_{m=1}^N$}
		\STATE Update $\bphi_{n\rightarrow m}$ and $\bphi_{m \rightarrow n }$ using \eqref{eq:update_phi_D1} and \eqref{eq:update_phi_D0}.		
		\ENDFOR		
		\STATE Update $\bpsi_n$ using \eqref{eq:update_psi}. 
		\STATE Update $\bgamma_n$ using \eqref{eq:update_gamma}.
		\STATE Update $\bxi_n$ using \eqref{eq:update_xi}. 
		\ENDFOR
		\STATE Update $\blambda$ using \eqref{eq:update_Gk} and \eqref{eq:update_Hk}.		
		\STATE Update $\bnu$ using \eqref{eq:update_nu}. 		 
		\STATE Update $\bmu$ using \eqref{eq:update_mu}.
		
		\RETURN $\bphi$, $\bpsi$, $\bgamma$, $\bxi$, $\blambda$, $\bnu$, and $\bmu$.	
	\end{algorithmic} 
\end{algorithm}

\subsection{Social network parameter \texorpdfstring{$\bbeta$}{beta}}\label{sec:beta}

From our Bayesian network model in \cref{fig:model}, we have the following: (i) For $(n,m)\neq(n',m')$, conditioned on $z_{n\rightarrow m},z_{m\rightarrow n},z_{n'\rightarrow m'},z_{m'\rightarrow n'},\beta_{k}$, $\bD(n,m)$ and $\bD(n',m')$ are independent. (ii) $\beta_k$ and $\bz$ are independent. Thus, for $k=1, \ldots,K_s$, the posterior distribution of $\beta_{k}$ is given by
\begin{dmath*}
p(\beta_{k} \mid \bD,\bz)\\
\propto {\prod_{(n,m):z_{n\rightarrow m}=z_{m\rightarrow n}=k}p(\bD(n,m) \mid  z_{n\rightarrow m},z_{m\rightarrow n},\beta_{k})p(\beta_{k})}
\propto \prod_{(n,m)}\beta_{k}^{\bD(n,m)I(z_{n \rightarrow m},k)I(z_{m \rightarrow n},k)}\cdot{(1-\beta_{k})}^{(1-\bD(n,m))I(z_{n \rightarrow m},k)I(z_{m \rightarrow n},k)}\Beta{g_0}{h_0}
\propto \beta_{k}^{\sum_{(n,m)}\bD(n,m)I(z_{n \rightarrow m},k)I(z_{m \rightarrow n},k)+g_0-1}\cdot{(1-\beta_{k})}^{\sum_{(n,m)}(1-\bD(n,m))I(z_{n \rightarrow m},k)I(z_{m \rightarrow n},k)+h_0-1}
\propto \calB e\left(\bmeta_k(\bD, \bz)\right),
\end{dmath*}
where
\begin{dmath*}
\bmeta_k(\bD, \bz)=\left({\sum_{(n,m)} \bD(n,m) I(z_{n \rightarrow m},k) I(z_{m \rightarrow n},k)+g_0},\\  \sum_{(n,m)}(1-\bD(n,m))I(z_{n \rightarrow m},k)I(z_{m \rightarrow n},k)+h_0\right).
\end{dmath*}

We choose the variational distribution of $\beta_k$ to be in the same exponential family as its posterior distribution. Let $\blambda_k=(G_{k}, H_{k})$ and the variational distribution of $\beta_k$ be $q(\beta_{k})=\Beta{G_k}{H_k}$. From \cite{BleKucMcA2017}, we obtain 
\begin{align*}
\blambda_k=\E{q(\bz)}[\bmeta_k(\bD, \bz)],
\end{align*}
with
\begin{align}
G_k&=\E_{q(\bz)}\left[\sum_{(n,m)}\bD(n,m)I(z_{n \rightarrow m},k)I(z_{m \rightarrow n},k)+g_0\right]\nonumber\\
&= \sum_{(n,m)}\bD(n,m)\phi_{n \rightarrow m,k}\phi_{m \rightarrow n,k}+g_0 \label{eq:update_Gk},\\
H_k&=\E_{q(\bz)}\left[\sum_{(n,m)}(1-\bD(n,m))I(z_{n \rightarrow m},k)I(z_{m \rightarrow n},k)+h_0\right]\nonumber\\
&= \sum_{(n,m)}(1-\bD(n,m))\phi_{n \rightarrow m,k}\phi_{m \rightarrow n,k}+h_0 \label{eq:update_Hk},
\end{align}
where $\phi_{n \rightarrow m,k}= q(z_{n \rightarrow m}=k)$ is defined in \cref{subsec:Community membershipz}.


A Beta distribution is a Dirichlet distribution with two parameters. Thus from equation (10) in \cite{Min2003}, we also have
\begin{align}
\E_{q(\beta_{k})}[\log(\beta_{k})]&=\Psi(G_k)-\Psi(G_k+H_k)\text{, and } \label{eq:Eq_logBeta} \\
\E_{q(\beta_{k})}[\log(1-\beta_{k})]&=\Psi(H_k)-\Psi(G_k+H_k),\label{eq:Eq_log1minBeta}
\end{align}
which are used in computing the variational distributions of other parameters in our model. Recall that $\Psi(\cdot)$ is the digamma function.

\subsection{Community membership indicators \texorpdfstring{$\bz$}{z}}\label{subsec:Community membershipz}

Consider two agents $n$ and $m$ with $\bD(n,m)=1$. From our Bayesian network model in \cref{fig:model}, for each $k={1, \ldots,K_s}$, we have 
\begin{dmath*}
{p(z_{n \rightarrow m}=k\mid \bpi_n,z_{m \rightarrow n},\bD(n,m)=1,\beta_k)}
\propto {p(\bD(n,m)=1\mid z_{n \rightarrow m}=k,\bpi_n,z_{m \rightarrow n},\beta_k)}\cdot{p(z_{n \rightarrow m}=k\mid \bpi_n,z_{m \rightarrow n},\beta_k)}
= {p(\bD(n,m)=1\mid z_{n \rightarrow m}=k,z_{m \rightarrow n},\beta_k)p(z_{n \rightarrow m}=k\mid \bpi_n)}
= \beta_{k}^{I(z_{m \rightarrow n},k)}\epsilon^{(1-I(z_{m \rightarrow n},k))}\pi_{n,k} \text{ .}
\label{eq:poster_z}
\end{dmath*}
Therefore, $p(z_{n \rightarrow m} \mid \bpi_n, z_{m \rightarrow n},\bD(n,m)=1, \beta_k)$ is a categorical distribution, which is an exponential family with natural parameter
\begin{dmath*}
\left(\log\left({\beta_{k}^{I(z_{m \rightarrow n},k)}\epsilon^{(1-I(z_{m \rightarrow n},k))}\pi_{n,k}}\right)-\log\left({\sum_{k=1}^{K_s}(\beta_{k}^{I(z_{m \rightarrow n},k)}\epsilon^{(1-I(z_{m \rightarrow n},k))}\pi_{n,k})}\right)\right)_{k=1}^{K_s}.
\end{dmath*}
We let the variational distribution of $z_{n->m}$ to be in the same exponential family as its posterior distribution, namely a categorical distribution. Assume its categorical probabilities are $(\phi_{n\rightarrow m,k})_{k=1}^{K_s}$. From \cite{BleKucMcA2017}, we obtain

\begin{dmath}
	\log{\phi_{n \rightarrow m,k}}
	=\E_{q(\beta_{k},\bpi_{n},z_{m \rightarrow n})}\left[\log\left({\beta_{k}^{I(z_{m \rightarrow n},k)}\epsilon^{(1-I(z_{m \rightarrow n},k))}\pi_{n,k}}\right)\right]-\E_{\prod_{k'=1}^{K_s}q(\beta_{k'},\bpi_n,z_{m \rightarrow n})}\left[\log\left(\sum_{k'=1}^{K_s}(\beta_{k'}^{I(z_{m \rightarrow n},k')} \\\cdot\epsilon^{(1-I(z_{m \rightarrow n},k'))}\pi_{n,k'})\right)\right].\label{eq:log_phi}
\end{dmath} 

The second term on the right-hand side of \eqref{eq:log_phi} is constant for every $\{\phi_{n\rightarrow m,k}\}_{k=1}^{K_s}$. Define  
\begin{align*}
&\Delta_{n\rightarrow m,k}\\
&\triangleq\exp\left\{\E{q(\beta_{k},\bpi_{n},z_{m \rightarrow n})}[\log(\beta_{k}^{I(z_{m \rightarrow n},k)}\epsilon^{(1-I(z_{m \rightarrow n},k))}\pi_{n,k})]\right\},
\end{align*} 
which is the exponential function of the first term on the right-hand side of \eqref{eq:log_phi}.
Then $\phi_{n \rightarrow m,k}=\Delta_{n\rightarrow m,k}\cdot c$, where $c$ is constant for every $\{\phi_{n\rightarrow m,k}\}_{k=1}^{K_s}$. Since $\sum_{k=1}^{K_s}\phi_{n\rightarrow m,k}=1$, when computing $\phi_{n\rightarrow m,k}$, we only need to compute $\Delta_{n\rightarrow m,k}$ to obtain $\phi_{n\rightarrow m,k}=\dfrac{\Delta_{n\rightarrow m,k}}{\sum_{k=1}^{K_s}\Delta_{n\rightarrow m,k}}$.
We have
\begin{dmath}
\Delta_{n\rightarrow m,k}
=\exp\left\{\phi_{m \rightarrow n,k}\E_{q(\beta_{k})}[\log(\beta_{k})]+(1-\phi_{m \rightarrow n,k})\log(\epsilon)+\E{q(\bpi_{n})}[\log(\pi_{n,k})]\right\}
\propto \exp\left\{\phi_{m \rightarrow n,k}\left(\E{q(\beta_{k})}[\log(\beta_{k})]-\log(\epsilon)\right)+\E_{q(\bpi_{n})}[\log(\pi_{n,k})]\right\}\label{eq:update_phi_D1},
\end{dmath}
where $\E_{q(\beta_{k})}[\log(\beta_{k})]$ and $\E_{q(\bpi_{n})}[\log(\pi_{n,k})]$ are computed using \eqref{eq:Eq_logBeta} and \eqref{eq:Eq_log_pi} in the sequel, respectively.  


Similarly, if $\bD(n,m)=0$, we have
\begin{align*}
&p(z_{n \rightarrow m}=k\mid \bpi_n,z_{m \rightarrow n},\bD(n,m)=0,\beta_k)\\
&\propto(1-\beta_{k})^{I(z_{m \rightarrow n},k)}(1-\epsilon)^{(1-I(z_{m \rightarrow n},k))}\pi_{n,k},
\end{align*}
and
\begin{dmath}
\Delta_{n\rightarrow m,k}
\triangleq\exp\{\E_{q(\beta_{k},\bpi_{n},z_{m \rightarrow n})}[\log((1-\beta_{k})^{I(z_{m \rightarrow n},k)}\\ \cdot(1-\epsilon)^{(1-I(z_{m \rightarrow n},k))}\pi_{n,k})]\}
=\exp\{\phi_{m \rightarrow n,k}\E_{q(\beta_{k})}[\log(1-\beta_{k})]+(1-\phi_{m \rightarrow n,k})\log(1-\epsilon)+\E_{q(\bpi_{n})}[\log(\pi_{n,k})]\}
\propto \exp\{\phi_{m \rightarrow n,k}\left(\E_{q(\beta_{k})}[\log(1-\beta_{k})]-\log(1-\epsilon)\right)+\E_{q(\bpi_{n})}[\log(\pi_{n,k})]\}\label{eq:update_phi_D0},
\end{dmath}
where $\E_{q(\beta_{k})}[\log(1-\beta_{k})]$ is computed in \eqref{eq:Eq_log1minBeta} in the sequel.


\subsection{Event community indices \texorpdfstring{$\bs$}{s}}\label{subsec:Community memberships}

From our Bayesian network model in \cref{fig:model}, for $k=1, \ldots,K_s$, the posterior distribution of $s_n^l$ is 
\begin{align*}
&p(s_n^l=k\mid \bpi_n, y_n^l,\theta^l,\bomega_{n,k})\\
&\propto p(y_n^l\mid s_n^l=k,\bpi_n,\theta^l,\bomega_{n,k})p(s_n^l=k\mid \bpi_n)\\
&=\bomega_{n,k}(\theta^l,y_n^l)\pi_{n,k}.
\end{align*}
Similar to Section \ref{subsec:Community membershipz}, we let the variational distribution of $s_n^{l}$ to be a categorical distribution with categorical probabilities $(\psi^l_{n,k})_{k=1}^{K_s}$. We then have
\begin{dmath}
\psi_{n,k}^l \propto \exp\{\E_{q(\bomega_{n,k}(\theta^l,\cdot),\theta^l,\bpi_n)}[\log(\bomega_{n,k}(\theta^l,y_n^l)\pi_{n,k})]\}
=\exp\left\{\E_{q(\theta^l)}\left[\E_{q(\bomega_{n,k}(\theta^l,\cdot))}[\log(\bomega_{n,k}(\theta^l,y_n^l))]\right]+\E_{q(\bpi_{n})}[\log(\pi_{n,k})]\right\}
=\exp\left\{\E_{q(\theta^l)}\left[\Psi(\bxi_{n,k}(\theta^l,y_n^l))-\Psi(\sum_{r'=1}^{R}\bxi_{n,k}(\theta^l,r'))\right]+\E_{q(\bpi_{n})}[\log(\pi_{n,k})]\right\}
=\exp\left\{\sum_{r=1}^{R}\bnu^l(r)\left[\Psi(\bxi_{n,k}(r,y_n^l))-\Psi(\sum_{r'=1}^{R}\bxi_{n,k}(r,r'))\right]+\E_{q(\bpi_{n})}[\log(\pi_{n,k})]\right\}\label{eq:update_psi},
\end{dmath}
where $\E_{q(\bomega_{n,k}(\theta^l,\cdot))}[\log(\bomega_{n,k}(\theta^l,y_n^l))]$ and $\E_{q(\bpi_{n})}[\log(\pi_{n,k})]$ are computed in \eqref{eq:Eq_log_omega} and \eqref{eq:Eq_log_pi}, respectively, and $\bnu^l(r)$ is defined in \eqref{eq:q_prior_theta}. 

\subsection{Community weights \texorpdfstring{$\bpi$}{pi}}\label{subsec:Community weights}
Since conditioned on $\bpi_n={(\pi_{n,k})_{k=1}^{K_s}}$, $\{s_n^l\}_{l=1}^L$ and $\{z_{n\rightarrow m}\}_{m=1,m\neq n}^N$ are independent, we have
\begin{align*}
&p\left(\bpi_n\mid \{s_n^l\}_{l=1}^L, \{z_{n\rightarrow m}\}_{m=1,m\neq n}^N \right)\\
&\propto \prod_{l=1}^{L}p(s_n^l\mid \bpi_n)\prod_{m=1,m\neq n}^{N}p(z_{n\rightarrow m}\mid \bpi_n)p(\bpi_n)\\
&\propto \Dir{\left(\frac{\alpha}{K_s}+\sum_{m=1,m\neq n}^{N}I(z_{n\rightarrow m},k)+\sum_{l=1}^{L}I(s_n^l,k)\right)_{k=1}^{K_s}}.
\end{align*}
We let the variational distribution of $\bpi_n$ to be in the same exponential family as its posterior distribution by letting $q(\bpi_n)\triangleq \Dir{\bgamma_n}$, where $\gamma_n$ is a vector of $K_s$ elements with $k$-th element being
\begin{dmath}
\gamma_{n,k}=\E_{q(\{s_n^l\}_{l=1}^L,\{z_{n\rightarrow m}\}_{m=1,m\neq n}^N)}\left[\frac{\alpha}{K_s}+\sum_{m=1,m\neq n}^{N}I(z_{n\rightarrow m},k)+\sum_{l=1}^{L}I(s_n^l,k)\right]
=\frac{\alpha}{K_s}+\sum_{m=1,m\neq n}^{N}\phi_{n\rightarrow m,k}+\sum_{l=1}^{L}\psi_{n,k}^l \label{eq:update_gamma}.
\end{dmath}

From (10) in \cite{Min2003}, we also have
\begin{align}
\E_{q(\bpi_{n})}[\log(\pi_{n,k})]=\Psi(\gamma_{n,k})-\Psi(\sum_{k=1}^{K_s}\gamma_{n,k}),\label{eq:Eq_log_pi}
\end{align}
which is used in \cref{subsec:Community membershipz,subsec:Community memberships}.

\subsection{Event states \texorpdfstring{$\btheta$}{theta}}\label{subsec:Event states}
For $r=1, \ldots,R$, the posterior distribution of $\theta^l$ is
\begin{align}
&p\left(\theta^l=r\mid \{y_n^l\}_{n=1}^{N},\bomega,\{s_n^l\}_{n=1}^{N}\right)\nonumber\\
&\propto \prod_{n=1}^{N}p\left(y_n^l\mid\theta^l=r,\{\bomega_{n,k}\}_{k=1}^{K_s},s_n^l\right)p(\theta^l=r).\label{eq:posterior_theta}
\end{align}
where $p(y_n^l\mid\theta^l=r,\{\bomega_{n,k}\}_{k=1}^{K_s},s_n^l) = \Cat{\bomega_{n,s_n^l}(r,\cdot)}$ and $\theta^l\sim \Unif{1,\ldots,R}$.
Thus,
\begin{align*}
p(\theta^l=r\mid \{y_n^l\}_{n=1}^{N},\bomega,\{s_n^l\}_{n=1}^{N})
\propto \prod_{n=1}^{N}\bomega_{n,s_n^l}(r,y_n^l).
\end{align*}
We let the variational distribution of $\theta^l$ be a categorical distribution, and denote
\begin{align}
q(\theta^l=r)\triangleq\bnu^l(r),\label{eq:q_prior_theta}
\end{align}
where
\begin{dmath}
\bnu^l(r)
\propto \exp\left\{\sum_{n=1}^{N}\E_{q(\bomega_{n,s_n^l}(r,\cdot),s_n^l)}\left[\log\left(\bomega_{n,s_n^l}(r,y_n^l)\right)\right]\right\}
=\exp\left\{\sum_{n=1}^{N}\E_{q(s_n^l)}\left[\E_{q(\bomega_{n,s_n^l}(r,\cdot)}\left[\log\left(\bomega_{n,k}(r,y_n^l)\right)\right]\right]\right\}
=\exp\left\{\sum_{n=1}^{N}\E_{q(s_n^l)}\left[\Psi(\bxi_{n,s_n^l}(r,y_n^l))-\Psi(\sum_{r'=1}^{R}\bxi_{n,s_n^l}(r,r'))\right]\right\}
=\exp\left\{\sum_{n=1}^{N}\sum_{k=1}^{K_s}\psi_{n,k}^l\left[\Psi(\bxi_{n,k}(r,y_n^l))-\Psi(\sum_{r'=1}^{R}\bxi_{n,k}(r,r'))\right]\right\}\label{eq:update_nu}.
\end{dmath}
The penultimate equation holds because
\begin{dmath*}
\E_{q(\bomega_{n,s_n^l}(r,\cdot)}\left[\log\left(\bomega_{n,k}(r,y_n^l)\right)\right]\\
= \Psi(\bxi_{n,k}(r,y_n^l))-\Psi(\sum_{r'=1}^{R}\bxi_{n,k}(r,r')),
\end{dmath*} 
which is the expectation of logarithm of $\bomega_{n,k}(r,y_n^l)$, the $(r,y_n^l)$-th element of the confusion matrix of agent $n$ when it is in community $k$ (see \eqref{eq:Eq_log_omega}).


\subsection{Agent confusion matrices \texorpdfstring{$\bomega$}{omega}}

The posterior distribution of $\bomega_{n,k}(r,\cdot)$ is
\begin{dmath*}
p(\bomega_{n,k}(r,\cdot)\mid \{y_n^l\}_{l=1}^{L},\{s_n^l\}_{l=1}^{L},\btheta,\tbomega_k(r,\cdot))
\propto\prod_{l\in\{1, \ldots,L\}:s_n^l=k,\theta^l=r} {p(y_n^l\mid \bomega_{n,k}(r,\cdot),s_n^l,\theta^l)}\cdot {p(\bomega_{n,k}(r,\cdot )\mid \tbomega_{k}(r,\cdot))}
=\prod_{l=1}^{L}\bomega_{n,k}(r,y_n^l)^{I(s_n^l,k)I(\theta^l,r)}\Dir{\tbomega_{k}(r,\cdot)}
= \Dir{\bzeta(r,r')_{r'=1}^R},
\end{dmath*}
where $\bzeta(r,r')=\tbomega_{k}(r,r')+\sum_{l=1}^{L}I(y_n^l,r')I(s_n^l,k)I(\theta^l,r)$.
Letting the variational distribution of $\bomega$ be the same as its posterior distribution, we have
\begin{align}
q(\bomega_{n,k}(r,\cdot))= \Dir{\bxi_{n,k}(r,\cdot)},\label{eq:VarEquation_xi}
\end{align}
where
$\bxi_{n,k}(r,\cdot)=(\bxi_{n,k}(r,r'))_{r'=1}^{R}$ and 
\begin{dmath}
\bxi_{n,k}(r,r')=\E_{q(\tbomega_{k}(r,\cdot),s_n^l,\theta^l)}\left[\tbomega_{k}(r,r')+\sum_{l=1}^{L}I(y_n^l,r')I(s_n^l,k)I(\theta^l,r)\right]
=\bmu_{k}(r,r')+\sum_{l=1}^{L}I(y_n^l,r')\psi_{n,k}^l\bnu^l(r)\label{eq:update_xi},
\end{dmath}
and $\bmu_{k}(r,r')$ is defined in \eqref{eq: q_tomega}.

Furthermore, we have
\begin{align}
&\E_{q(\bomega_{n,k})}[\log(\bomega_{n,k}(r,r'))]\nonumber\\
&=\Psi(\bxi_{n,k}(r,r'))-\Psi(\sum_{r^{''}=1}^{R}\bxi_{n,k}(r,r^{''}))\label{eq:Eq_log_omega},
\end{align}
which is used in \cref{subsec:Community memberships} and \cref{subsec:Event states}.

\subsection{Community confusion matrices \texorpdfstring{$\tbomega$}{omega}}
The prior distribution of $\tbomega_{k}(r,\cdot)$ and its likelihood in our model are not conjugate, thus we cannot use standard mean-field variational inference to update its variational distribution $q(\tbomega_{k}(r,\cdot))$. In order to find $q(\tbomega_{k}(r,\cdot))$ that maximizes \eqref{eq:CostFunction}, we take the functional derivative of the objective function \eqref{eq:CostFunction} with respect to $q(\tbomega_{k}(r,\cdot))$ and set it to zero, namely $\frac{\partial \calL(q)}{\partial q(\tbomega_{k}(r,\cdot))}=0$ to obtain the maximizer as
\begin{dmath}
	q(\tbomega_{k}(r,\cdot))
	\propto \exp\left\{\E_{ \prod_{n=1}^{N}q(\bomega_{n,k}(r,\cdot))} \left[{\log p(\{\bomega_{n,k}(r,\cdot)\}_{n=1}^N\mid\tbomega_k(r,\cdot) )}+\log p(\tbomega_k(r,\cdot))\right]\right\}	
	=\exp\left\{\sum_{n=1}^{N}\left\{\log\Gamma(\sum_{r'=1}^{R}\tbomega_{k}(r,r'))-\sum_{r'=1}^{R}\log\Gamma(\tbomega_{k}(r,r'))
	+\sum_{r'=1}^{R}(\tbomega_{k}(r,r')-1)[\Psi(\bxi_{n,k}(r,r'))-\sum_{r'=1}^{r}\Psi(\bxi_{n,k}(r,r'))]\right\}
	-{\frac{R}{2}\log(2\pi)-\frac{1}{2}\log(\det(\bV))}-\sum_{r'=1}^{R}\log(\tbomega_{k}(r,r'))-\frac{1}{2}(\log(\tbomega_{k}(r,\cdot))-\bM)^{T}\bV^{-1}(\log(\tbomega_{k}(r,\cdot))-\bM)\right\}. \label{eq: objectForOmegak}
\end{dmath}

Equation \eqref{eq: objectForOmegak} is difficult to analyze directly. Thus we use Laplace approximation method introduced in \cite{Bis2006, WanBle2013} to find a Gaussian approximation to it, namely
\begin{align}
q(\tbomega_{k}(r,\cdot))\approx \calN\left(\bmu_{k}(r,\cdot),-\frac{1}{\nabla^2\log\left(q(\bmu_{k}(r,\cdot))\right)}\right),\label{eq: q_tomega}
\end{align}
where $\bmu_{k}(r,\cdot)$ is given by
\begin{align}
\bmu_{k}(r,\cdot)=\argmax_{\tbomega_k(r,\cdot)}\ \log\left(q(\tbomega_{k}(r,\cdot))\right)\label{eq:update_mu},
\end{align}
which can be solved using the gradient descent algorithm. 

\section{Three-level Stochastic Variational Inference Method}\label{sec:StoVarInfer}
The traditional variational inference method uses the whole dataset in each iteration and thus it does not scale well to large datasets.
To mitigate this, we propose the use of stochastic optimization and a three-level Stochastic VISIT (S-VISIT) algorithm to update parameters using randomly sampled subsets. The pseudo code of our top-level algorithm is shown in Algorithm \ref{alg:inferenceAlgorithm}.

The variational parameters corresponding to the variables in our model in \cref{fig:model} can be divided into three levels as follows:
\begin{enumerate}[(i)]
	\item Agent pair level: The variational parameters $(\phi_{n\to m})_{n\ne m}$ corresponding to the community membership indicators $(z_{n\to m})_{n\ne m}$ provide information about the social relationships between agents.
	\item Agent level: The variational parameters $\bpsi_n$, $\bgamma_n$, and $(\bxi_{n,k})_k$ correspond to agent $n$'s event community indices $\{s_n^l\}_l$, community weights $\bpi_n$, and confusion matrices $(\bomega_{n,k})_k$.  
	\item Social network and event level: The variational parameters $\{\blambda_k\}_k$ and $\{\bnu^l\}_l$ corresponding to the social network parameters $\{\bbeta_k\}_k$ and event states $\btheta$. 
\end{enumerate}

\begin{algorithm}
	\caption{S-VISIT ($i$-th iteration)}\label{alg:inferenceAlgorithm}
	\begin{algorithmic}[H]
		\renewcommand{\algorithmicrequire}{\textbf{Input:}}
		\renewcommand{\algorithmicensure}{\textbf{Output:}}
		\REQUIRE Variational parameters in $(i-1)$-th iteration, opinions $\by$, social network data $\bD$.
		\ENSURE  Variational parameters in $i$-th iteration. 		
		\STATE Sample a subset $\calS_n$ from agent set $\{1, \ldots, N\}$.		
		\STATE\textbf{Social network and event level start:}
		\FOR {each agent n in $\calS_n$}
		\STATE\textbf{Agent level start:}
		\STATE Sample a subset $\calS_p$ from pair set $\{(n,m)_{m=1,m\neq n}^N\}$.
		\FOR {each agent pair (n,m) in $\calS_p$}
		\STATE\textbf{Agent pair level start:}
		\STATE Update $\bphi_{n\rightarrow m}$ and $\bphi_{m\rightarrow n}$ using \eqref{eq:update_phi_D1} and \eqref{eq:update_phi_D0}.
		\STATE\textbf{Agent pair level end.}		
		\ENDFOR		
		\STATE Update $\bpsi_n$ using \eqref{eq:update_psi}. 
		\STATE Update $\bgamma_n$ using \eqref{eq:update_gamma_sto}.
		\STATE Update $\bxi_n=(\bxi_{n,k})_k$ using \eqref{eq:update_xi}. 
		\STATE\textbf{Agent level end.}
		\ENDFOR
		\STATE Update $\blambda$ using \eqref{eq:update_lambda_sto}.		
		\STATE Updata $\bnu$ using \eqref{eq:update_nu_sto}. 		 
		\STATE Update $\bmu$ using \eqref{eq:update_mu_sto}.
		\STATE\textbf{Social network and event level end.}
		\RETURN $\bphi$, $\bpsi$, $\bgamma$, $\bxi$, $\blambda$, $\bnu$, and $\bmu$.	
	\end{algorithmic} 
\end{algorithm}

Consider $\gamma_{n,k}$, the variational parameter of $\pi_{n,k}$, as an example.  We fix the rest of variational parameters and calculate the natural gradient\cite{Ama1998} of \eqref{eq:CostFunction} with respect to $\gamma_{n,k}$:
\begin{align}
\nabla_{\gamma_{n,k}}\calL(q)=\frac{\alpha}{K_s}+\sum_{m=1,m\neq n}^{N}\phi_{n\rightarrow m,k}+\sum_{l=1}^{L}\psi_{n,k}^l-\gamma_{n,k}.\label{eq:NatureGradient_gamma}
\end{align}
Let $\gamma_{n,k}\tc{i}$ be the update of $\gamma_{n,k}$ at the $i$-th iteration. If we let $\nabla_{\gamma_{n,k}}\calL(q)=0$ and update $\gamma_{n,k}$, we need to use the whole set of $\{\phi_{n\rightarrow m,k}\}_{m=1,m\neq n}^N$. Instead of using \eqref{eq:NatureGradient_gamma} to update $\gamma_{n,k}$, we can use stochastic approximation to randomly sample a subset $\calS_p$ from the set $\{(n,m)_{m=1,m\neq n}^N\}$, and update $\gamma_{n,k}$ by using
\begin{align}
\gamma_{n,k}^{(i)}=\gamma_{n,k}^{(i-1)}+\rho^{(i)}(\hgamma_{n,k}^{(i)}-\gamma_{n,k}^{(i-1)}),\label{eq:update_gamma_sto}
\end{align}
where
\begin{align}
\hgamma_{n,k}^{(i)}
=\frac{\alpha}{K_s}+\dfrac{N}{|\calS_p|}\sum_{(n,m)\in\calS_p}\phi_{n\rightarrow m,k}^{(i)}+\sum_{l=1}^{L}\psi_{n,k}^{l,(i)}\label{eq:update_gamma_sto_subequaton}
\end{align}
and $\rho^{(i)}$ is the step size at $i$-th iteration.

In \eqref{eq:update_gamma_sto_subequaton} we only use the randomly sampled subset $\{\phi_{n\rightarrow m,k}\}_{(n,m)\in\calS_p}$ instead of the whole set $\{\phi_{n\rightarrow m,k}\}_{m=1,m\neq n}^{N}$ to perform the update. If the sequence of step sizes $(\rho^{(i)})_i$ of all the iterations satisfies 
\begin{align}
\sum_{i=1}^{\infty}\rho^{(i)}=\infty \text{ and } \sum_{i=1}^{\infty}(\rho^{(i)})^2<\infty,\label{eq:convergeCondition}
\end{align}
then from \cite{RobMon1951}, $\gamma_{n,k}\tc{i}$ converges to a local optimum.

Similarly, we can update the other two variational parameters $\blambda_k$ and $\bnu^l(r)$, which correspond to the global parameters $\beta_k$ and $\theta^l$ respectively. At each iteration, $|\calS_n|$ agents are randomly selected from the agent set $\{1, \ldots,N\}$. For $r=1, \ldots,R$, we have the following equations:
\begin{align}
G_{k}^{(i)}&=G_{k}^{(i-1)}+\rho^{(i)}(\widehat{G}_{k}^{(i)}-G_{k}^{(i-1)}),\nonumber\\
H_{k}^{(i)}&=H_{k}^{(i-1)}+\rho^{(i)}(\widehat{H}_{k}^{(i)}-H_{k}^{(i-1)}),\nonumber\\
\blambda_k^{(i)} &=[G_{k}^{(i)}, H_{k}^{(i)}]^T,\label{eq:update_lambda_sto}\\
\bnu^{l,(i)}(r)&=\bnu^{l,(i-1)}(r)+\rho^{(i)}(\widehat{\bnu}^{l,(i)}(r)-\bnu^{l,(i-1)}(r)),\label{eq:update_nu_sto}
\end{align}
where
\begin{dmath*}
\widehat{G}_{k}^{(i)}
= \dfrac{N}{|\calS_n|}\dfrac{N}{|\calS_p|}\sum_{n\in\calS_n,(n,m)\in\calS_p}\bD(n,m)\phi^{(i)}_{n \rightarrow m,k}\phi^{(i)}_{m \rightarrow n,k}+g_0,\\
\widehat{H}_{k}^{(i)}
=\dfrac{N}{|\calS_n|}\dfrac{N}{|\calS_p|} \sum_{n\in\calS_n,(n,m)\in\calS_p}(1-\bD(n,m))\phi^{(i)}_{n \rightarrow m,k}\phi^{(i)}_{m \rightarrow n,k}+h_0,\\
{\widehat{\bnu}^{l,(i)}(r)\propto}\exp\left\{\dfrac{N}{\left|\calS_n\right|}\sum_{n\in\calS_n}\sum_{k=1}^{K_s}\psi_{n,k}^{l,(i)}\left(\Psi(\bxi^{(i)}_{n,k}(r,y_n^l)) \\- \Psi(\sum_{r'=1}^{R}\bxi^{(i)}_{n,k}(r,r'))\right)\right\}.
\end{dmath*}

When updating $\bmu^{(i)}_{k}(r,\cdot)$, the variational parameter of $\tbomega_{k}(r,\cdot)$, instead of using \eqref{eq: objectForOmegak}, we use its noisy but unbiased estimator, which is given by
\begin{dmath}
	\widehat{q}^{(i)}(\tbomega_{k}(r,\cdot))	
	=\exp\left\{\dfrac{N}{\left|\calS_n\right|}\sum_{n\in\calS_n}\left\{\log\Gamma(\sum_{r'=1}^{R}\tbomega_{k}(r,r'))-\sum_{r'=1}^{R}\log\Gamma(\tbomega_{k}(r,r'))
	+\sum_{r'=1}^{R}(\tbomega_{k}(r,r')-1)[\Psi(\bxi^{(i)}_{n,k}(r,r'))-\sum_{r'=1}^{r}\Psi(\bxi^{(i)}_{n,k}(r,r'))]\right\}
	-{\frac{R}{2}\log(2\pi)-\frac{1}{2}\log(\det(\bV))}-\sum_{r'=1}^{R}\log(\tbomega_{k}(r,r'))-\frac{1}{2}(\log(\tbomega_{k}(r,\cdot))-\bM)^{T}\bV^{-1}(\log(\tbomega_{k}(r,\cdot))-\bM)\right\} \label{eq: objectForOmegakSto}
\end{dmath}
then we update
\begin{align}
\bmu_{k}^{(i)}(r,\cdot)=\argmax_{\tbomega_k(r,\cdot)}\ \log[\widehat{q}^{(i)}(\tbomega_{k}(r,\cdot))] \label{eq:update_mu_sto},
\end{align}
We use the gradient descent algorithm to find $\bmu_{k}^{(i)}(r,\cdot)$ in \cref{eq:update_mu_sto}.
At each iteration, multiple gradient descent steps are conducted using the sub-dataset to update
$\bmu_{k}^{(i)}(r,\cdot)$, which is then set as the initial value in the next iteration.




\section{Simulation and Experiment Results}\label{sec:SimulationResults}

In this section, we present simulations and real data experiments to evaluate VISIT and S-VISIT methods. For comparison, we adopt three state-of-the-art methods as the baseline methods, namely majority voting, TruthFinder, AccuSim, CATD, BCC, and CBCC. We demonstrate that our methods outperform the baseline methods for inferring event states as well as confusion matrices when agents either remain in the same community or switch communities when observing different events.

\subsection{Agents remain in the same community when observing different events}
As both BCC and CBCC require each agent have a unique confusion matrix when observing different events, we first consider the scenario where agents remain in the same community for all the events. 

\subsubsection{Synthetic Data Generation}\label{sec: SynDataGen}\label{sec:Agents_remain_in_same_community}
The event states are selected from a set of $R=6$ states. We use both opinions from agents and the social network to infer the event states. 
We set the number of agents, events, and communities to $N=80$, $L=200$, and $K=4$, respectively. We set $\beta_k=0.9$ for $k=1,\ldots,K$. Note that $K$ is used here to generate the synthetic dataset and is smaller than the maximum number of communities $K_s$ we set in our inference method. For agents $1:20$, $21:40$, $41:60$ and $61:80$, we set $\bpi_n$ to be  $(\tfrac{9}{10},\tfrac{1}{30},\tfrac{1}{30},\tfrac{1}{30})$, $(\tfrac{1}{30},\tfrac{9}{10},\tfrac{1}{30},\tfrac{1}{30})$, $(\tfrac{1}{30},\tfrac{1}{30},\tfrac{9}{10},\tfrac{1}{30})$, and $(0,\tfrac{3}{10},\tfrac{7}{10},0)$, respectively.
We sample $z_{n\rightarrow m}, s_n, \bD(n,m)$, and $ y_n^l$ from \eqref{eq:prior_znm}, \eqref{eq:prior_s}, \eqref{eq:prior_Dnm}, and \eqref{eq:Observation}, respectively. 

It is worth emphasizing that we keep agents in the same community for all the events as required by both BCC and CBCC. Thus, we let $s_n^1=\ldots=s_n^L$ and sample their common value from a categorical distribution with parameter $\bpi_n$, which means the confusion matrix of each agent remains the same across different events. Let the confusion matrix $\bomega_{k}$ have the same value $d_k$ on its diagonal and every non-diagonal element be $\dfrac{1-d_k}{R-1}$, where $d_k$ equals 0.05, 0.1, 0.5, 0.2 for $k=1, \ldots,4$. Let $\bA$ be an $N\times L$ observation matrix, whose $(i,j)$-th element $\bA(i,j)$ is equal to $1$ if agent $i$ observes event $j$, and $\bA(i,j)$ is equal to $0$ otherwise. Let $\aleph$ be the proportion of zero elements in $\bA$. We call $\aleph$ the \emph{sparsity} of $\bA$. We generate synthetic datasets for 5 sparsity values, which are 0.7, 0.75, 0.8, 0.85, and 0.9, respectively.

\subsubsection{Truth discovery accuracy}\label{section:AS_NC}
In the simulation, we set the maximum number of communities $K_s$ to be 10. We set hyper-parameters $g_0, h_0, \bM,\bV$ and $\alpha$ to be $1,1,(2,2,2,2,2,2),0.7\calI$ and $0.1$ respectively, where $\calI$ is a $6\times6$ identity matrix. For different sparsity values, we generate different data sets and conducted 50 Monte Carlo experiments for each method and each sparsity value of the observation matrices.  We define the accuracy score as the number of correct estimated events, divided by the total number of events and then averaged over all Monte Carlo experiments. 

\begin{figure}[!htb]
	\centering
	\includegraphics[width=8cm]{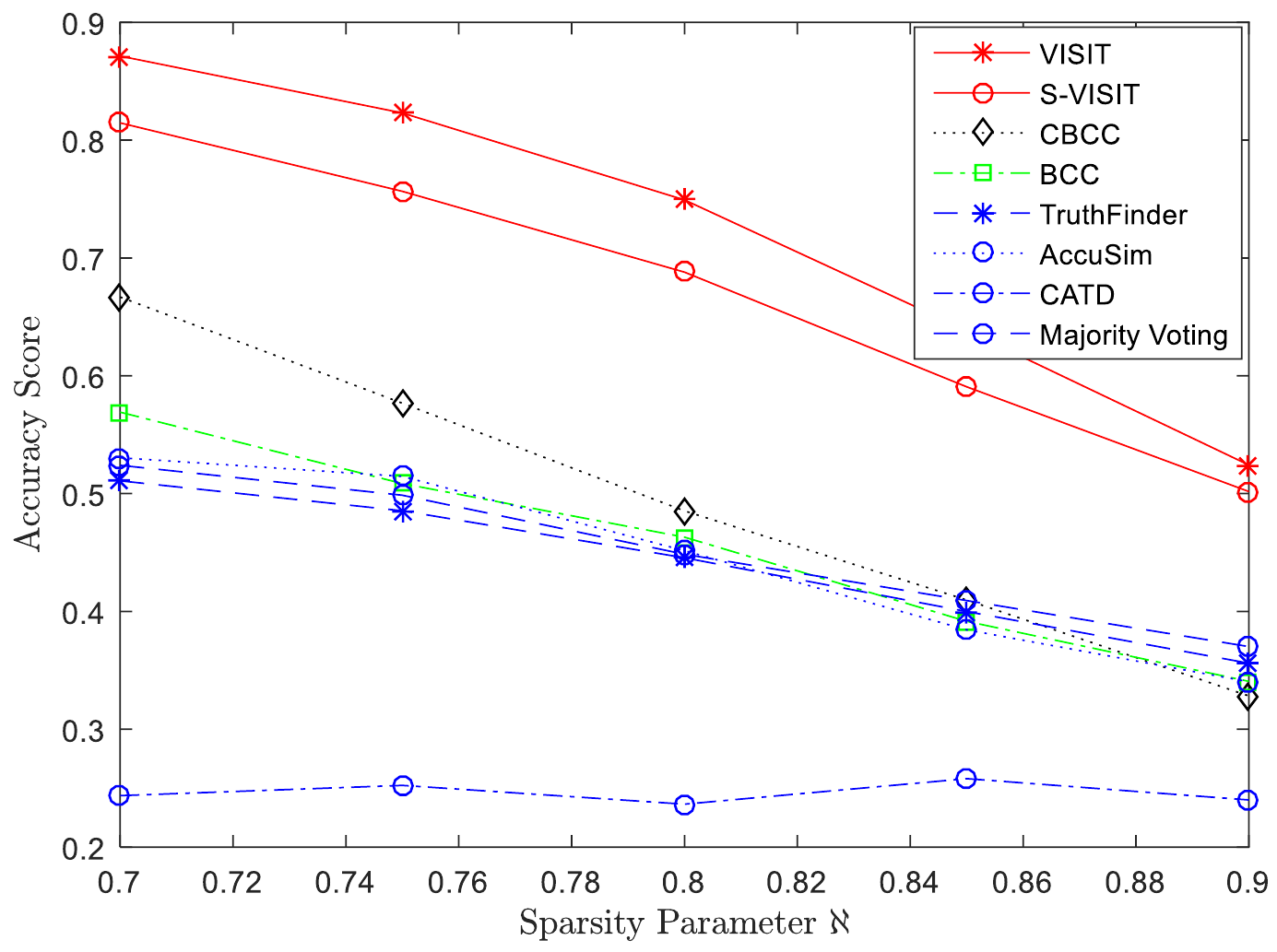}
	\caption{Accuracy scores when agents remain in the same community.}
	\label{fig:AC_NC}	
\end{figure}

The result is shown in \cref{fig:AC_NC}. We observe that BCC, CBCC, TruthFinder, AccuSim, CATD, VISIT and S-VISIT have better performance than the majority voting method because these methods model the reliabilities or the confusion matrices of agents and can better explain the quality
of observations. VISIT, S-VISIT and CBCC have better performance than BCC because they take into account the community of the agents and can thus better estimate the confusion matrix when the observation matrix is sparse. VISIT outperforms CBCC by almost 20\% because VISIT uses information about the social network, which improves the clustering performance. The accuracy score of S-VISIT is lower than VISIT, as expected. It is noteworthy that when the observation matrix is very sparse, the performance of all the methods can be worse than majority voting. 

\subsubsection{Estimation of the confusion matrices}

\begin{figure}[!htb]
	\centering
	\includegraphics[width=8cm]{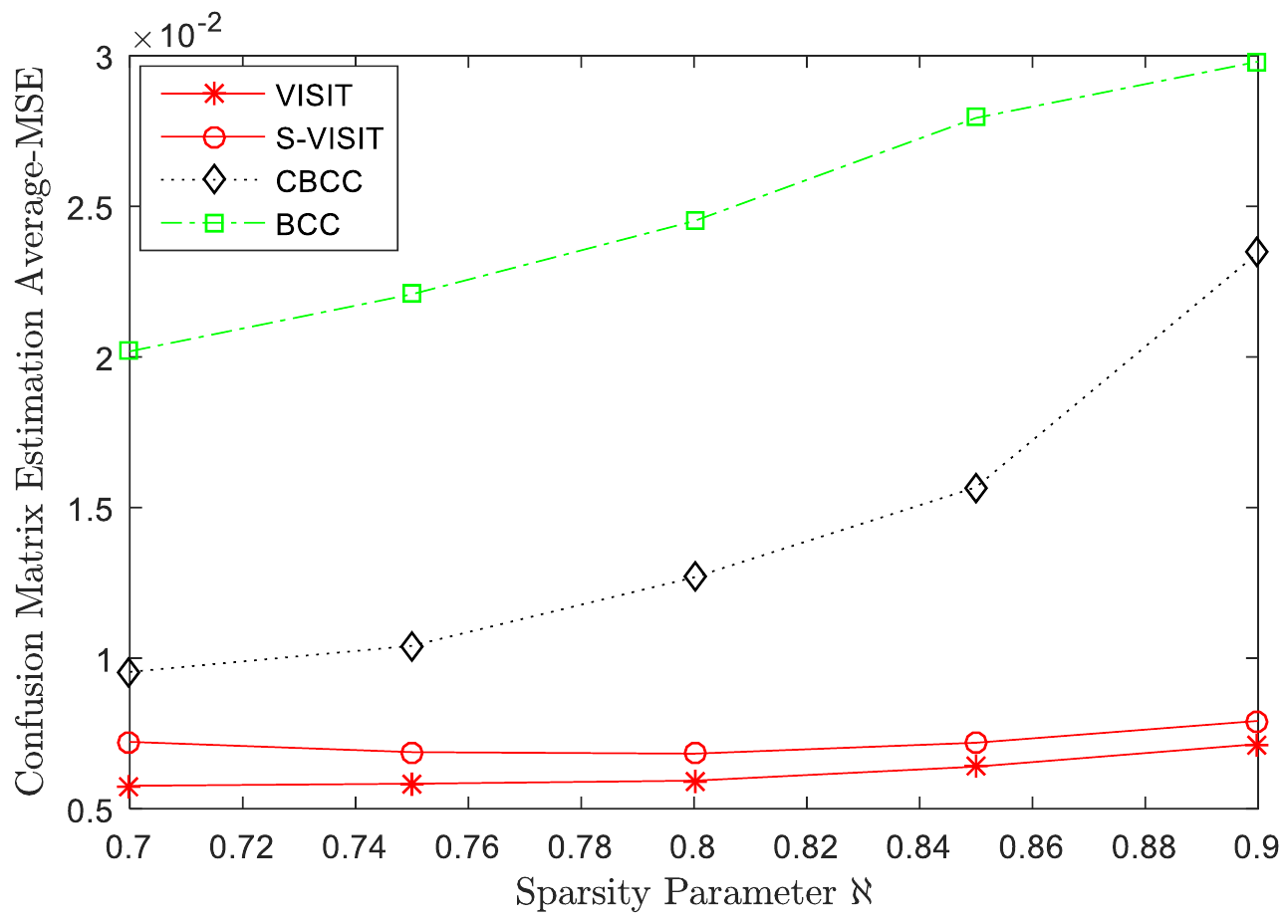}
	\caption{Confusion matrix MSEs when agents remain in the same community.}
	\label{fig:CM_NC}	
\end{figure}

Let $\bomega^{*}_{n}$ be the true confusion matrix of agent $n$. The mean square error (MSE) in estimating $\bomega^{*}_{n}$ is defined as:
\begin{align}
\text{MSE} = \frac{\sum_{n=1}^N\sum_{l=1}^L\sum_{r=1}^R \sum_{r'=1}^R |\bomega_{n,s_n^l}(r,r') - \bomega^{*}_{n}(r,r')|^2}{N\cdot L\cdot R \cdot R}.\label{eq: MSEDefinition_Our}
\end{align}
For CBCC and BCC, as the community indices remain the same when observing different events, we denote the estimated confusion matrix as $\bomega_n'$ and calculate the MSE as 
\begin{align}
\text{MSE} = \frac{\sum_{n=1}^N\sum_{r=1}^R \sum_{r'=1}^R |\bomega_{n}'(r,r') - \bomega^{*}_{n} (r,r')|^2}{N \cdot R \cdot R}.\label{eq: MSEDefinition_BCC&CBCC}
\end{align}
We then take the average of the MSEs over 50 Monte Carlo experiments. As TruthFinder, AccuSim and CATD do not use confusion matrices, we do not consider them in this comparison. The result is shown in \cref{fig:CM_NC}.  Our methods have better performance than BCC and CBCC. When the observation matrix is sparse, our methods and  CBCC collect the observations in the same community to estimate the community confusion matrix, which is then used to estimate confusion matrices for the agents belonging to that community. This yields better results than BCC, which does not assume any community structure. Compared with CBCC, our methods are better able to estimate the communities because we use social network information to aid in the procedure.

\subsection{Agents switch communities when observing different events}

In some applications, agents may subscribe to the beliefs of different communities when observing different events. In this subsection, we evaluate the performance of our proposed methods when agents switch communities when observing different events. 

\subsubsection{Synthetic Data Generation}
We use the same settings as \cref{sec:Agents_remain_in_same_community} except when sampling the community indices. In \cref{sec:Agents_remain_in_same_community}, we let $s_n^1=\ldots=s_n^L$ and sample their common value from a categorical distribution with parameter $\bpi_n$. In this subsection, we sample each $s_n^l$ independently from $\bpi_n$. Therefore, the confusion matrices of an agent may vary across different events.

\subsubsection{Truth discovery accuracy}
\begin{figure}[!htb]
	\centering
	\includegraphics[width=8cm]{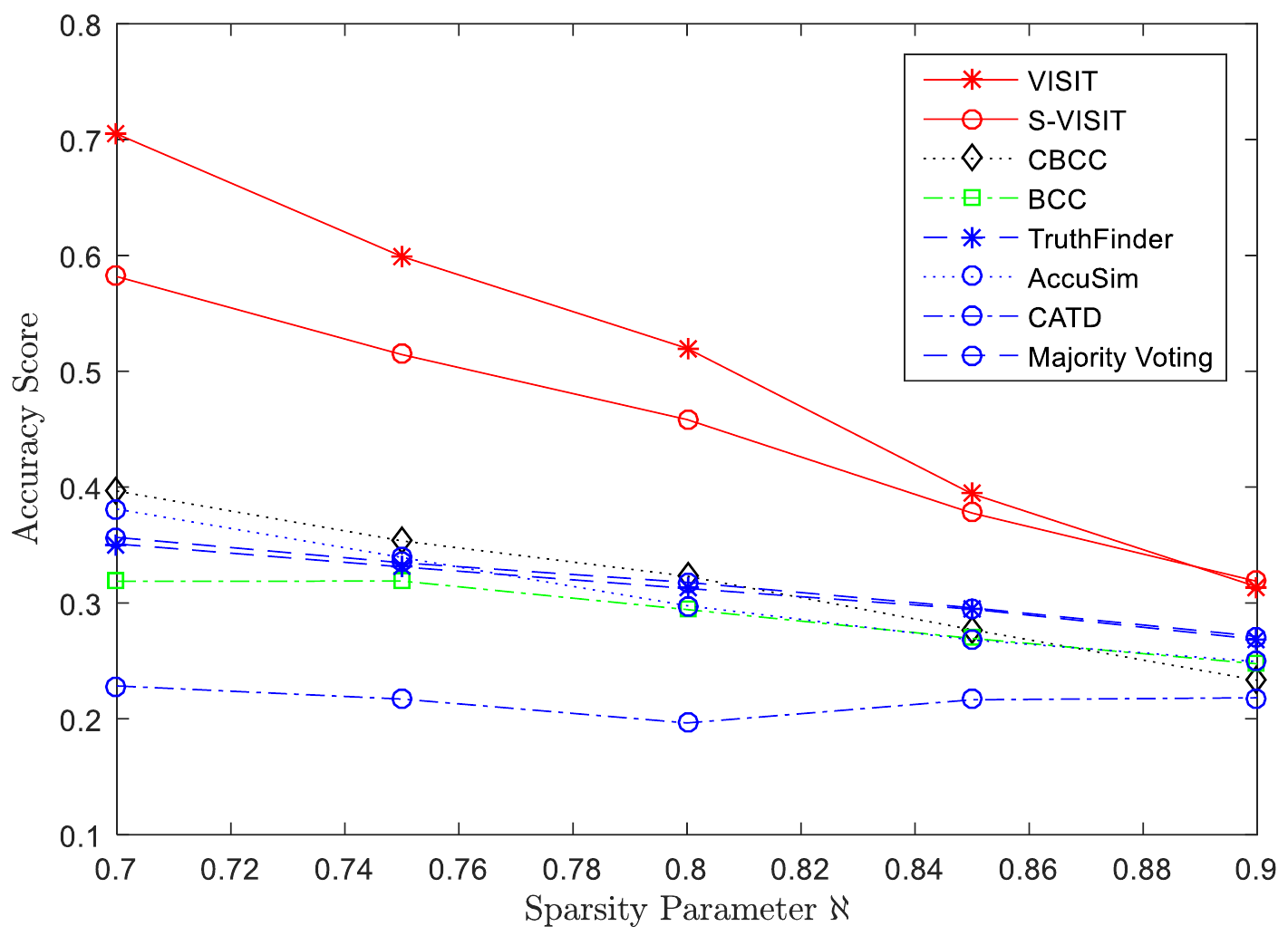}
	\caption{Accuracy scores when agents switch communities.}
	\label{fig:AC_C}	
\end{figure}

From \cref{fig:AC_C}, we observe that the performance of S-VISIT and VISIT are better than the six baseline methods. This is because our methods account for the agents switching community beliefs over different events, while CBCC assume that agents do not switch communities. The other baseline methods do not consider agent communities.

\subsubsection{Estimation of the confusion matrices}
\begin{figure}[!htb]
	\centering
	\includegraphics[width=8cm]{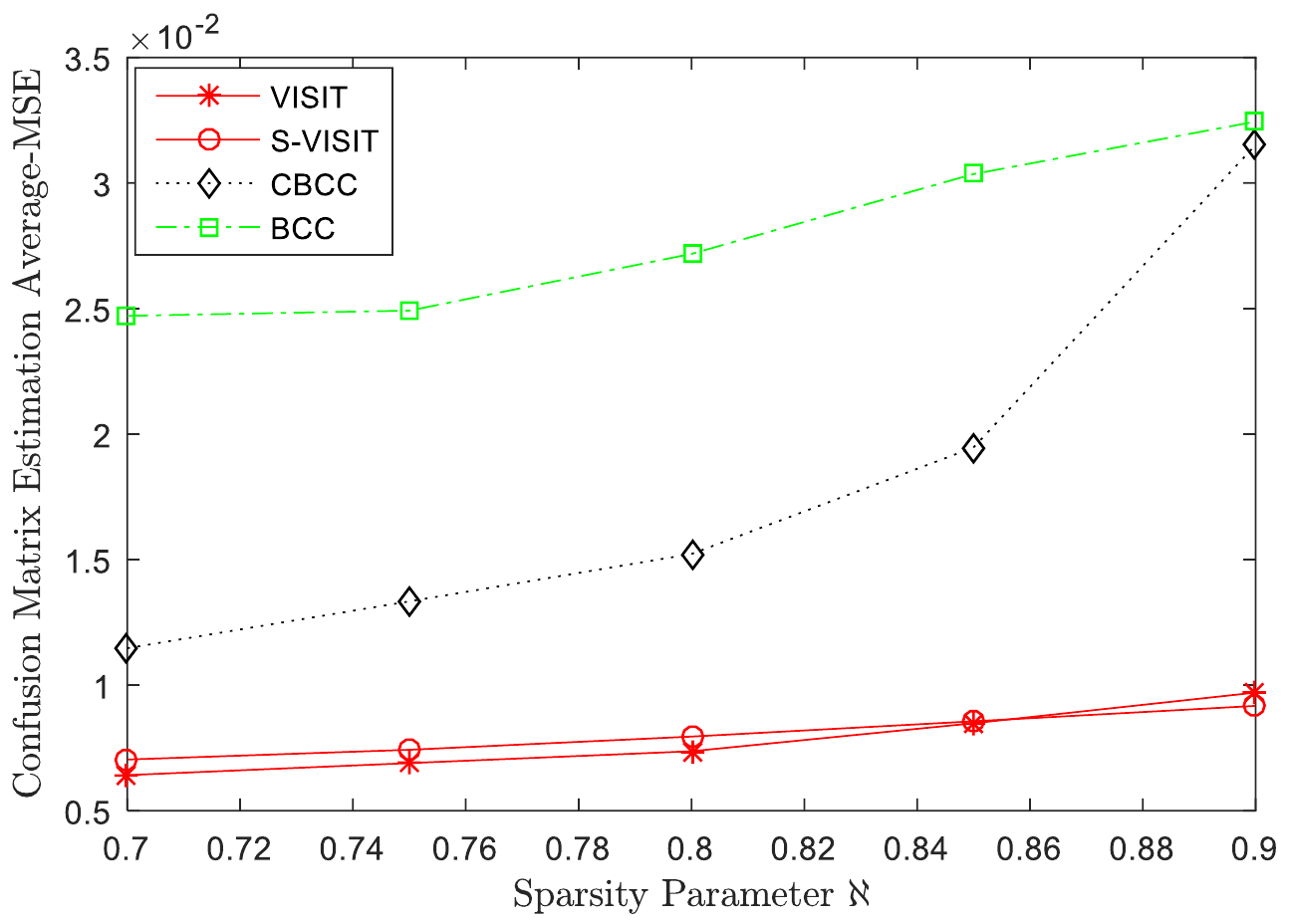}
	\caption{Confusion matrix MSEs when agents switch communities.}
	\label{fig:CM_C}	
\end{figure}

From \cref{fig:CM_C}, we observe that VISIT method performs the best, and the performance of our S-VISIT method is also better than the three baseline methods.

\subsection{Real data}
We next compare the performance of our proposed methods on two real datasets against the three baseline methods. 
\subsubsection{Movie ranking dataset}

\begin{figure}[!htb]
	\centering
	\includegraphics[width=7cm]{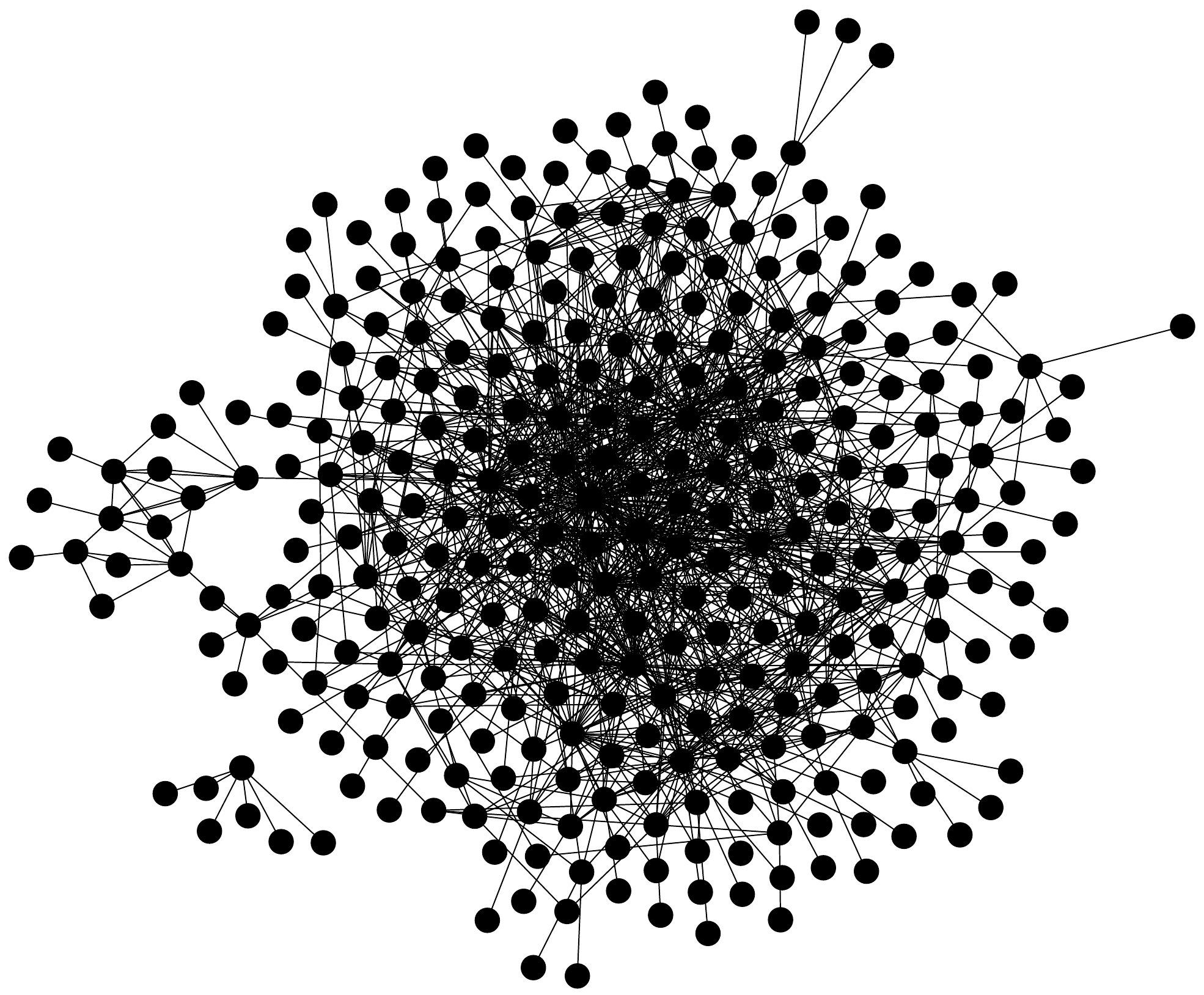}
	\caption{The social network of users in IMDB movie ranking}.	
	\label{Fig:NET_TOPOLOGY_REAL}
\end{figure}

The real dataset consists of movie evaluations from IMDB,\footnote{http://www.imdb.com/} which provides a platform where individuals can evaluate movies on a scale of 1 to 10. If a user rates a movie and clicks the share button, a Twitter message is generated. We then extract the rating from the Twitter message. In our first experiment, we divide the movie evaluations into 4 levels: bad (0-4), moderate (5,6), good (7,8), and excellent (9,10). In our second experiment, we divide the movie evaluations into 2 levels: bad (0-5), good (6-10). We treat the ratings on the IMDB website as the event truths, which are based on the aggregated evaluations from all users, whereas our observations come from only a subset of users who share their ratings on Twitter. Using the Twitter API, we collect information about the follower and following relationships between individuals that generate movie evaluation Twitter messages. To better show the influence of social network information on event truth discovery, we delete small subnetworks that consist of less than 5 agents. The final dataset \cite{YanTay2018} we use consists of 2266 evaluations from 209 individuals on 245 movies (events) and also the social network between these 209 individuals. Similar to \cite{For2010,GopBle2013}, we regard the social network to be undirected as both follower or following relationships indicate that the two users have similar taste. The social network is shown in \cref{Fig:NET_TOPOLOGY_REAL}.

The performance of different methods are shown in Table \ref{table:Acc_RealData1}.  We observe that our proposed methods perform better than the other benchmark methods in both experiments.

\begin{table}[!htb]
	\caption{Truth discovery accuracy on movie ranking dataset.} 
	\centering 
	\begin{tabular}{c c c} 
		\hline 
		Method & 4 Levels & 2 Levels \\ [0.5ex]
		\hline		
		Majority Voting  & 0.50 & 0.71 \\ [0.5ex]
		TruthFinder      & 0.52 & 0.71 \\ [0.5ex]
		AccuSim          & 0.53 & 0.70 \\ [0.5ex]
		CATD             & 0.49 & 0.70 \\ [0.5ex]
		BCC              & 0.51 & 0.69 \\ [0.5ex]
		CBCC             & 0.53 & 0.71 \\ [0.5ex]
		S-VISIT          & 0.55 & 0.73 \\ [0.5ex]
		VISIT            & 0.55 & 0.75 \\ [0.5ex]		
		\hline 
	\end{tabular}
	\label{table:Acc_RealData1} 
\end{table}

\subsubsection{Russian president rumor dataset}
The dataset from \cite{YaoHuLi2016} contains rumors regarding the Russian president Vladimir V. Putin, who had not appeared in public for more than one week up to 14 March 2015. Various speculations including false reports were posted on Twitter. The Twitter following relationships are provided in the dataset. We select the Twitter users with more than two followers to form the social network. As a Twitter user only reports an event to be true if he believes that it has occurred, to enhance the complexity of the dataset, for each event we randomly choose 20\% of users in the social network who did not post any report about that event, and set their observations about the event to be opposite of the ground truth. The performance of the different methods are shown in \cref{table:Acc_RealData2}.  We observe that VISIT has the best performance, followed by CBCC.

\begin{table}[!ht]
	\caption{Truth discovery accuracy on Russian president rumor dataset.} 
	\centering 
	\begin{tabular}{l l} 
		\hline 
		Method & Accuracy \\ [0.5ex]
		\hline 	
		Majority Voting & 0.985    \\ [0.5ex]
		TruthFinder     & 0.985    \\ [0.5ex]
		AccuSim         & 0.985    \\ [0.5ex]
		CATD            & 0.985    \\ [0.5ex]
		BCC             & 0.985    \\ [0.5ex]
		CBCC            & 0.991    \\ [0.5ex]
		S-VISIT         & 0.985    \\ [0.5ex]
		VISIT           & 0.992    \\ [0.5ex] 
	
		\hline 
	
	\end{tabular}
	\label{table:Acc_RealData2} 
\end{table}

\section{Conclusion}\label{sec:Conclusion}
In this paper, we have proposed a truth discovery method based on social network communities. Similar to other community based methods, our method performs better than methods without considering communities when the observation matrix is sparse. We incorporate information about the social network into the truth discovery framework to improve the truth discovery performance and accuracy of estimating the agents' confusion matrices. We have developed a Laplace variational method and a three-level stochastic variational inference method to infer our non-conjugate model, with simulation and experimental results suggesting that the performance of our proposed approaches are better than several other inference methods, including majority voting, TruthFinder, AccuSim, CATD, BCC and CBCC methods. Unlike the other methods, in our model, each agent can subscribe to the beliefs of different communities when observing different events.


%




\ifCLASSOPTIONcaptionsoff
  \newpage
\fi



%



\bibliographystyle{IEEEtran}
\bibliography{IEEEabrv,TruthDiscovery}

\begin{thebibliography}{10}
\providecommand{\url}[1]{#1}
\csname url@samestyle\endcsname
\providecommand{\newblock}{\relax}
\providecommand{\bibinfo}[2]{#2}
\providecommand{\BIBentrySTDinterwordspacing}{\spaceskip=0pt\relax}
\providecommand{\BIBentryALTinterwordstretchfactor}{4}
\providecommand{\BIBentryALTinterwordspacing}{\spaceskip=\fontdimen2\font plus
\BIBentryALTinterwordstretchfactor\fontdimen3\font minus
  \fontdimen4\font\relax}
\providecommand{\BIBforeignlanguage}[2]{{%
\expandafter\ifx\csname l@#1\endcsname\relax
\typeout{** WARNING: IEEEtran.bst: No hyphenation pattern has been}%
\typeout{** loaded for the language `#1'. Using the pattern for}%
\typeout{** the default language instead.}%
\else
\language=\csname l@#1\endcsname
\fi
#2}}
\providecommand{\BIBdecl}{\relax}
\BIBdecl

\bibitem{karger2013efficient}
D.~R. Karger, S.~Oh, and D.~Shah, ``Efficient crowdsourcing for multi-class
  labeling,'' \emph{ACM SIGMETRICS Performance Evaluation Review}, vol.~41,
  no.~1, pp. 81--92, 2013.

\bibitem{KanTay:C17}
Q.~Kang and W.~P. Tay, ``Sequential multi-class labeling in crowdsourcing: A
  {Ulam-Renyi} game approach,'' in \emph{IEEE/WIC/ACM Int. Conf. on Web
  Intelligence}, Leipzig, Germany, Aug. 2017.

\bibitem{AceDahLobOzd:11}
D.~Acemoglu, M.~A. Dahleh, I.~Lobel, and A.~Ozdaglar, ``Bayesian learning in
  social networks,'' \emph{Review of Economic Studies}, vol.~78, no.~4, pp.
  1201--1236, Mar. 2011.

\bibitem{HoTayQue:J15}
J.~Ho, W.~P. Tay, T.~Q.~S. Quek, and E.~K.~P. Chong, ``Robust decentralized
  detection and social learning in tandem networks,'' \emph{{IEEE} Trans.
  Signal Process.}, vol.~63, no.~19, pp. 5019 -- 5032, Oct. 2015.

\bibitem{Tay:J15}
W.~P. Tay, ``Whose opinion to follow in multihypothesis social learning? {A}
  large deviations perspective,'' \emph{IEEE J. Sel. Topics Signal Process.},
  vol.~9, no.~2, pp. 344 -- 359, Mar. 2015.

\bibitem{GraSurAli2016}
M.~L. Gray, S.~Suri, S.~S. Ali, and D.~Kulkarni, ``The crowd is a collaborative
  network,'' in \emph{ACM Conf. Computer-Supported Cooperative Work \& Social
  Computing}, 2016, pp. 134--147.

\bibitem{HuaWan2016}
C.~Huang and D.~Wang, ``Topic-aware social sensing with arbitrary source
  dependency graphs,'' in \emph{ACM/IEEE Int. Conf. Inform. Process. Sensor
  Networks}, 2016, pp. 1--12.

\bibitem{KanTay:J18}
Q.~Kang and W.~P. Tay, ``Sequential {Multi-Class} labeling in crowdsourcing,''
  \emph{IEEE Trans. Knowl. Data Eng.}, 2018.

\bibitem{LiGaoMen2015}
Y.~Li, J.~Gao, C.~Meng, Q.~Li, L.~Su, B.~Zhao, W.~Fan, and J.~Han, ``A survey
  on truth discovery,'' \emph{arXiv:1505.02463}, 2015.

\bibitem{MarWan2016}
J.~Marshall and D.~Wang, ``Mood-sensitive truth discovery for reliable
  recommendation systems in social sensing,'' in \emph{Proc. ACM Conf.
  Recommender Syst.}, 2016, pp. 167--174.

\bibitem{GarXioSun2017}
D.~A. Garcia-Ulloa, L.~Xiong, and V.~Sunderam, ``Truth discovery for
  spatio-temporal events from crowdsourced data,'' \emph{Proceedings of the
  {VLDB} Endowment}, vol.~10, no.~11, pp. 1562--1573, 2017.

\bibitem{WanKapLe2012}
D.~Wang, L.~Kaplan, H.~Le, and T.~Abdelzaher, ``On truth discovery in social
  sensing: {A} maximum likelihood estimation approach,'' in \emph{ACM/IEEE Int.
  Conf. Inform. Process. Sensor Networks}, 2012, pp. 233--244.

\bibitem{WanAmiLi2014}
D.~Wang, M.~T. Amin, S.~Li, T.~Abdelzaher, L.~Kaplan, S.~Gu, C.~Pan, H.~Liu,
  C.~C. Aggarwal, and R.~Ganti, ``Using humans as sensors: {an}
  estimation-theoretic perspective,'' in \emph{ACM/IEEE Int. Conf. Inform.
  Process. Sensor Networks}, 2014, pp. 35--46.

\bibitem{YaoHuLi2016}
S.~Yao, S.~Hu, S.~Li, Y.~Zhao, L.~Su, L.~Kaplan, A.~Yener, and T.~Abdelzaher,
  ``On source dependency models for reliable social sensing: {Algorithms} and
  fundamental error bounds,'' in \emph{IEEE Int. Conf. Distributed Computing
  Syst.}, 2016, pp. 467--476.

\bibitem{WanKapAbd2013}
D.~Wang, L.~Kaplan, T.~Abdelzaher, and C.~C. Aggarwal, ``On credibility
  estimation tradeoffs in assured social sensing,'' \emph{IEEE J. Sel. Areas
  Commun.}, vol.~31, no.~6, pp. 1026--1037, 2013.

\bibitem{MaTayXia:J18}
L.~Ma, W.~P. Tay, and G.~Xiao, ``Iterative expectation maximization for
  reliable social sensing with information flows,'' \emph{Information
  Sciences}, 2018.

\bibitem{YinHanPhi2008}
X.~Yin, J.~Han, and S.~Y. Philip, ``Truth discovery with multiple conflicting
  information providers on the web,'' \emph{IEEE Trans. Knowl. Data Eng.},
  vol.~20, no.~6, pp. 796--808, 2008.

\bibitem{DonBerSri2009}
X.~L. Dong, L.~Berti-Equille, and D.~Srivastava, ``Integrating conflicting
  data: the role of source dependence,'' \emph{Proc. VLDB Endowment}, vol.~2,
  no.~1, pp. 550--561, 2009.

\bibitem{LiLiGao2015a}
Q.~Li, Y.~Li, J.~Gao, L.~Su, B.~Zhao, M.~Demirbas, W.~Fan, and J.~Han, ``A
  confidence-aware approach for truth discovery on long-tail data,''
  \emph{Proc. VLDB Endowment}, vol.~8, no.~4, pp. 425--436, 2015.

\bibitem{ZhaWanZha2017}
D.~Y. Zhang, D.~Wang, and Y.~Zhang, ``Constraint-aware dynamic truth discovery
  in big data social media sensing,'' in \emph{IEEE Int. Conf. Big Data}, 2017,
  pp. 57--66.

\bibitem{ZhaRubGem2012}
B.~Zhao, B.~I.~P. Rubinstein, J.~Gemmell, and J.~Han, ``A {Bayesian} approach
  to discovering truth from conflicting sources for data integration,''
  \emph{Proc. VLDB Endowment}, vol.~5, no.~6, pp. 550--561, 2012.

\bibitem{KimGha2012}
H.~C. Kim and Z.~Ghahramani, ``Bayesian classifier combination,'' in
  \emph{Artificial Intell. and Stat.}, 2012, pp. 619--627.

\bibitem{ZheLiLi2017}
Y.~Zheng, G.~Li, Y.~Li, C.~Shan, and R.~Cheng, ``Truth inference in
  crowdsourcing: is the problem solved?'' \emph{Proc. VLDB Endowment}, vol.~10,
  no.~5, pp. 541--552, 2017.

\bibitem{VenGuiKaz2014}
M.~Venanzi, J.~Guiver, G.~Kazai, P.~Kohli, and M.~Shokouhi, ``Community-based
  {Bayesian} aggregation models for crowdsourcing,'' in \emph{Int. Conf. World
  Wide Web}, 2014, pp. 155--164.

\bibitem{YinGraSur2016}
M.~Yin, M.~L. Gray, S.~Suri, and J.~W. Vaughan, ``The communication network
  within the crowd,'' in \emph{Int. Conf. World Wide Web}, 2016, pp.
  1293--1303.

\bibitem{ZhaWuHua2017}
X.~Zhang, Y.~Wu, L.~Huang, H.~Ji, and G.~Cao, ``Expertise-aware truth analysis
  and task allocation in mobile crowdsourcing,'' in \emph{IEEE Int. Conf.
  Distributed Computing Syst.}, 2017, pp. 922--932.

\bibitem{KolFri2009}
D.~Koller and N.~Friedman, \emph{Probabilistic Graphical Models: Principles and
  Techniques (Adaptive Computation and Machine Learning series)}.\hskip 1em
  plus 0.5em minus 0.4em\relax The MIT Press, 2009.

\bibitem{AirBleFie2008}
E.~M. Airoldi, D.~M. Blei, S.~E. Fienberg, and E.~P. Xing, ``Mixed membership
  stochastic blockmodels,'' \emph{J. Mach. Learning Research}, vol.~9, no. Sep,
  pp. 1981--2014, 2008.

\bibitem{ForHri2016}
S.~Fortunato and D.~Hric, ``Community detection in networks: {A} user guide,''
  \emph{Physics Rep.}, vol. 659, pp. 1--44, 2016.

\bibitem{TehJorBea2012}
Y.~W. Teh, M.~I. Jordan, M.~J. Beal, and D.~M. Blei, ``Hierarchical {Dirichlet}
  processes,'' \emph{J. Am. Stat. Assoc.}, 2012.

\bibitem{IshZar2002}
H.~Ishwaran and M.~Zarepour, ``Exact and approximate sum representations for
  the {Dirichlet} process,'' \emph{Canadian J. Stat.}, vol.~30, no.~2, pp.
  269--283, 2002.

\bibitem{FoxSudJor2011}
E.~B. Fox, E.~B. Sudderth, M.~I. Jordan, and A.~S. Willsky, ``A sticky
  {HDP-HMM} with application to speaker diarization,'' \emph{Ann. Appl. Stat.},
  pp. 1020--1056, 2011.

\bibitem{HofBleWan2013}
M.~D. Hoffman, D.~M. Blei, C.~Wang, and J.~Paisley, ``Stochastic variational
  inference,'' \emph{J. Mach. Learning Research}, vol.~14, no.~1, pp.
  1303--1347, 2013.

\bibitem{BleKucMcA2017}
D.~M. Blei, A.~Kucukelbir, and J.~D. McAuliffe, ``Variational inference: A
  review for statisticians,'' \emph{J. Amer. Stat. Assoc.}, 2017.

\bibitem{WanBle2013}
C.~Wang and D.~M. Blei, ``Variational inference in nonconjugate models,''
  \emph{J. Mach. Learning Research}, vol.~14, no. Apr, pp. 1005--1031, 2013.

\bibitem{TieKasKad1989}
L.~Tierney, R.~E. Kass, and J.~B. Kadane, ``Fully exponential {Laplace}
  approximations to expectations and variances of nonpositive functions,''
  \emph{J. Amer. Stat. Assoc.}, vol.~84, no. 407, pp. 710--716, 1989.

\bibitem{Min2003}
\BIBentryALTinterwordspacing
T.~Minka, \emph{Bayesian inference, entropy, and the multinomial distribution},
  2003. [Online]. Available:
  \url{https://tminka.github.io/papers/minka-multinomial.pdf}
\BIBentrySTDinterwordspacing

\bibitem{Bis2006}
C.~M. Bishop, \emph{Pattern Recognition and Machine Learning}.\hskip 1em plus
  0.5em minus 0.4em\relax Springer, 2006.

\bibitem{Ama1998}
S.~I. Amari, ``Natural gradient works efficiently in learning,'' \emph{Neural
  Computation}, vol.~10, no.~2, pp. 251--276, 1998.

\bibitem{RobMon1951}
H.~Robbins and S.~Monro, ``A stochastic approximation method,'' \emph{Ann.
  Math. Stat.}, pp. 400--407, 1951.

\bibitem{YanTay2018}
\BIBentryALTinterwordspacing
J.~Yang and W.~P. Tay. (2018) Using social network information to discover
  truth of movie ranking. [Online]. Available:
  \url{https://doi.org/10.21979/N9/L5TTRW}
\BIBentrySTDinterwordspacing

\bibitem{For2010}
S.~Fortunato, ``Community detection in graphs,'' \emph{Physics Rep.}, vol. 486,
  no.~3, pp. 75--174, 2010.

\bibitem{GopBle2013}
P.~K. Gopalan and D.~M. Blei, ``Efficient discovery of overlapping communities
  in massive networks,'' \emph{Proc. Nat. Academy Sci.}, vol. 110, no.~36, pp.
  14\,534--14\,539, 2013.

\end{thebibliography}

%
\begin{IEEEbiography}[{\includegraphics[width=1in,height =1.25in,clip,keepaspectratio]{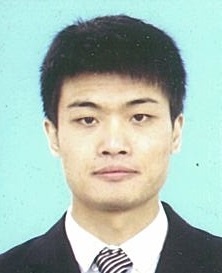}}]{Jielong Yang}
	received the B.Eng. and M.Sc. degree in Mechanical Engineering from Xi'an Jiaotong University in 2012 and 2014, respectively. He is currently a Ph.D. candidate at the School of Electrical and Electronic Engineering, Nanyang Technological University. His research interests include Bayesian network and Deep Neural Network.	
\end{IEEEbiography}
\begin{IEEEbiography}[{\includegraphics[width=1in,height =1.25in,clip,keepaspectratio]{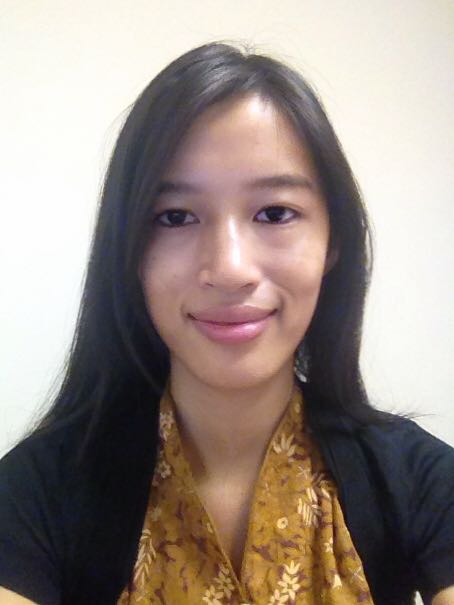}}]{Junshan Wang} 
received the B.S. and Ph.D. degrees in Statistics from Wuhan University and National University of Singapre (NUS), CN and SG, in 2011 and 2015, respectively. She spent two years in NUS as an instructor and worked as a research fellow in 2017 at Nanyang Technological University, Singapore. Currently she is a data scientist at Grab. Her recent interests are real world applications of Machine Learning and Deep Learning algorithms and techniques, in detecting fraud activity and risk control.
\end{IEEEbiography}
\begin{IEEEbiography}[{\includegraphics[width=1in,height =1.25in,clip,keepaspectratio]{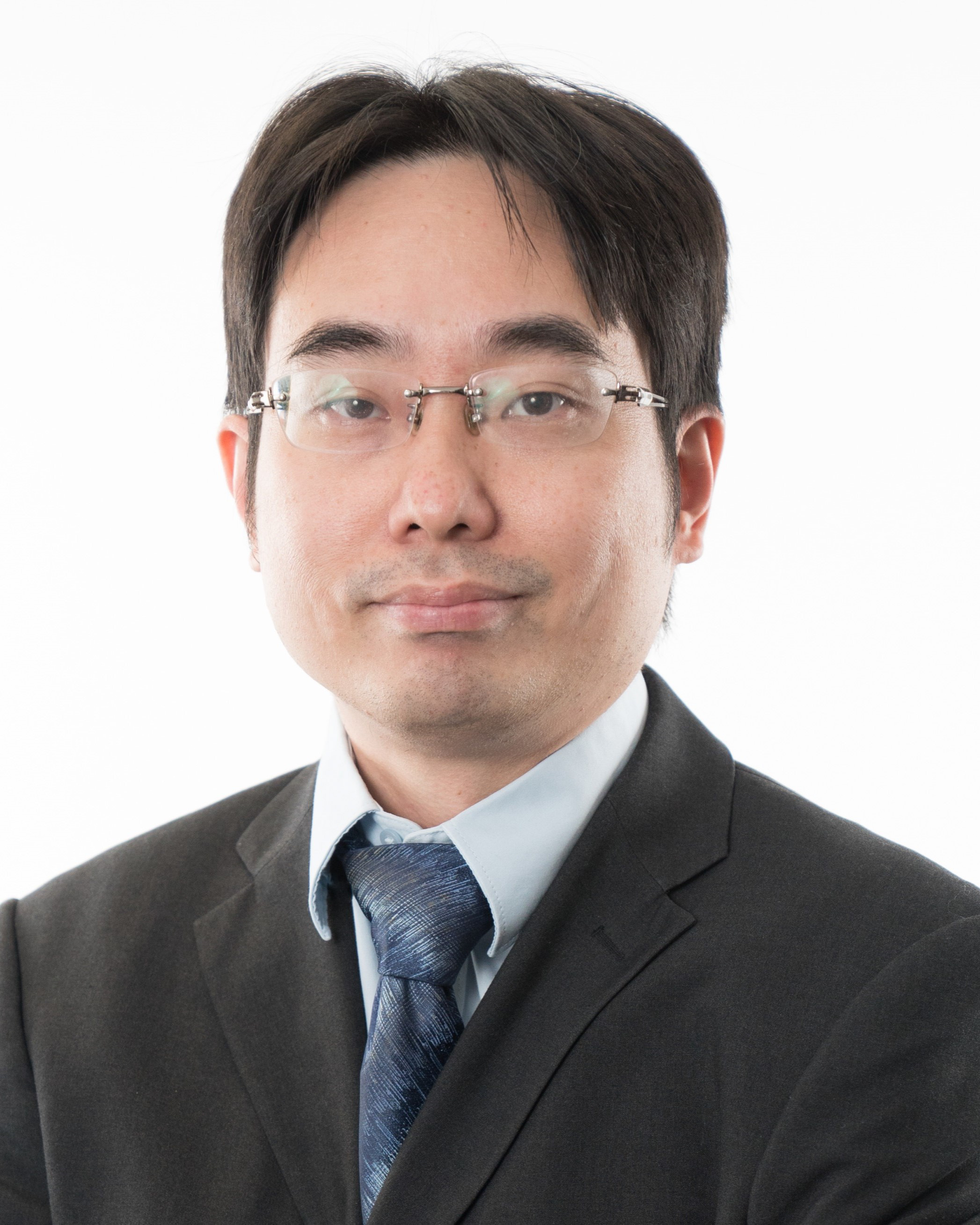}}]{Wee Peng Tay}(S'06 M'08 SM'14) received the B.S. degree in Electrical Engineering and Mathematics, and the M.S. degree in Electrical Engineering from Stanford University, Stanford, CA, USA, in 2002. He received the Ph.D. degree in Electrical Engineering and Computer Science from the Massachusetts Institute of Technology, Cambridge, MA, USA, in 2008. He is currently an Associate Professor in the School of Electrical and Electronic Engineering at Nanyang Technological University, Singapore. His research interests include information and signal processing over networks, distributed inference and estimation, information privacy, machine learning, information theory, and applied probability. Dr. Tay received the Tan Chin Tuan Exchange Fellowship in 2015. He is a coauthor of the best student paper award at the Asilomar conference on Signals, Systems, and Computers in 2012, and the IEEE Signal Processing Society Young Author Best Paper Award in 2016. He is currently an Associate Editor for the IEEE Transactions on Signal Processing, an Editor for the IEEE Transactions on Wireless Communications, and serves on the MLSP TC of the IEEE Signal Processing Society. He has also served as a technical program committee member for various international conferences. 
\end{IEEEbiography}

\end{document}